\documentclass[twocolumn, aps, pre]{revtex4-2}
\usepackage{graphicx}
\usepackage{multirow}
\usepackage{amsmath,amssymb,amsfonts}
\usepackage{amsthm}
\usepackage{mathrsfs}
\usepackage[title]{appendix}
\usepackage{xcolor}
\usepackage[hidelinks, bookmarks=false]{hyperref}
\hypersetup{
    colorlinks,
    linkcolor={blue!80!black},
    citecolor={red!80!black},
    urlcolor={blue!80!black}
}

\usepackage{textcomp}
\usepackage{booktabs}
\usepackage{algorithm}
\usepackage{algorithmicx}
\usepackage{listings}
\usepackage[T1]{fontenc}
\usepackage{tikz}
\usetikzlibrary{arrows.meta}
\usepackage{pgfplots}
\pgfplotsset{compat=1.17}
\usetikzlibrary{trees}
\usetikzlibrary{arrows.meta}
\newcommand{\cQ}{\mathcal{Q}}

\begin{document}

%\title{Power laws and hyperuniformity: Role of anisotropy and center-of-mass conservation}

\title{Power laws, anisotropy and center-of-mass conservation in mass transport processes}

%% Author Group
\author{Aniket Samanta}
\email{aniket.samanta@bose.res.in}
%\affiliation{Department of Physics of Complex Systems, S. N. Bose National Centre for Basic Sciences, Block-JD, Sector-III, Salt Lake, Kolkata 700106, India.}
\author{Animesh Hazra}
\email{ahazra.edu@gmail.com}
 %\affiliation{Department of Physics of Complex Systems, S. N. Bose National Centre for Basic Sciences, Block-JD, Sector-III, Salt Lake, Kolkata 700106, India.}
\author{Punyabrata Pradhan}
 %\email{punyabrata.pradhan@bose.res.in}
 \affiliation{Department of Physics of Complex Systems, S. N. Bose National Centre for Basic Sciences, Block-JD, Sector-III, Salt Lake, Kolkata 700106, India.}

\begin{abstract}
We present exact results for steady-state density correlation functions in conserved-mass transport processes with  {\it anisotropic}, reflection-symmetric hopping on a $d-$dimensional hypercubic lattice. 
In addition to mass conservation, we consider center-of-mass (CoM) conservation, imposed either along a specific axis or along all axes. CoM-conserving dynamics is implemented through coordinated {\it multidirectional} hopping of two equal chunks of masses in {\it opposite} directions. 
While anisotropy and mass conservation are known to generate power-law density correlations $C({\bf x}) \sim 1/|{\bf x}|^d$ at large distance $|{\bf x}| \gg 1$ {\it [Phys. Rev. A {\bf 42}, 1954 (1990)]}, an additional CoM conservation can qualitatively alter the nature of the power law. Indeed, when CoM is conserved in {\it all} directions, the correlations decay faster $-$ typically as $C({\bf x}) \sim 1/|{\bf x}|^{(d+2)}$, regardless of the presence (or absence) of anisotropy. Consequently, the systems exhibit an extreme {\it hyperuniformity} (``class I''), where the long-wavelength density fluctuations, despite the slow power-law decay, are anomalously suppressed. When CoM is conserved along particular ({\it not} all) directions, the slower $1/|{\bf x}|^{d}$ power-law decay is recovered.
The above behavior can be understood from an analogy between the correlation function and an electrostatic potential: While a (rank-$2$) quadrupolar charge distribution gives rise to the $1/|{\bf x}|^{d}$ power law, the $1/|{\bf x}|^{(d+2)}$ power law originates from a higher-order (rank-$4$) multipolar charge distribution. 
These findings reveal a rich interplay between anisotropy and CoM conservation in nonequilibrium steady states.
\end{abstract}

%\keywords{Power laws, Anisotropy, Mass and center-of-mass conservation, Hyperuniformity}

\maketitle

\section{Introduction}\label{sec1}

Power-law correlations are abundant in nature; they are observed in a wide variety of systems in a nonequilibrium setting, ranging from mountain ranges, river networks, and coastlines to the statistics of energy release in earthquakes and the ubiquitous $1/f$ noise in electrical conductors \cite{mandelbrot1982fractal, kirkby1983fractal, Scheidegger1967Mar, Bak2002Apr, Dutta1981Jul, Bak1996}. These systems exhibit scale invariance and fractal-like structures, reflected in spatio-temporal correlations that decay algebraically over wide ranges of length and time scales. A striking feature of such behavior is that power laws in driven systems often emerge without any apparent fine-tuning of external parameters \cite{Bak1987Jul}.

This generic emergence of algebraic correlations stands in sharp contrast to equilibrium systems with short-ranged interactions. In equilibrium, systems obey time-reversal symmetry, which leads to detailed balance, and consequently they are described by the Boltzmann-Gibbs distribution. The condition of detailed balance constrains the transition rates such that the fluctuation-dissipation theorem holds and power-law behavior is restricted to critical points that require precise tuning of control parameters, as exemplified by the critical point of the Ising lattice gas in the temperature-magnetic-field plane. In nonequilibrium, the situation is however qualitatively different, as detailed balance is violated and the steady-state probability weights $-$ unlike in equilibrium $-$ are generally non-Gibbsian, thus fundamentally altering the structure of correlations. In fact, the violation of detailed balance $-$ equivalently, the breaking of time-reversal symmetry $-$ is widely regarded as a necessary condition for the appearance of long-range correlations \cite{Procaccia1979Jan}, but it should be noted that the violation alone is not sufficient to result in power-law scaling \cite{Grinstein1990Apr, Garrido1990Aug, vanBeijeren1990Sep, Dickman1998May}. There are additional factors, such as symmetries, conservation laws, and dimensionality, which determine both the existence of algebraic decay and the associated scaling exponents. Indeed, despite significant progress \cite{Spohn1983Dec, Bertini2001Jul, Bertini2002May, Derrida2025May, Spohn2012Dec, Bertini2015Jun}, the problem of characterizing power laws in driven systems remain open and an actively pursued subject.

In this context, mass transport processes and the associated lattice models play a vital role, owing to their analytical tractability and computational efficiency, by providing a minimal and particularly useful framework for addressing these issues theoretically. Such models have been regularly employed to understand on a qualitative level a wide variety of physical phenomena, including cloud and gel formation, droplet and island growth, force transmission in granular materials, and avalanche dynamics, etc. Considerable progress has been made in the past in elucidating the dynamical origin of power laws in current-carrying systems, either driven by boundary reservoirs \cite{Procaccia1979Jan} or by localized disorder \cite{Sadhu2014Jul}. In these cases, the mechanisms leading to long-range correlations and scale-invariant behavior are now relatively well understood.
In contrast, isolated systems with strictly conserved mass and no net particle currents $-$ such as those arising from symmetric hopping (diffusive) dynamics $-$ are somewhat less understood.

For translationally invariant systems with mass-conserving diffusive dynamics, anisotropy has been identified as an important ingredient for generating long-range density correlation function $C({\bf x}) \sim 1/r^d$ having a power-law decay at large distance $r=|{\bf x}|$ in $d$ dimensions \cite{Grinstein1990Apr, Garrido1990Aug, Maes1990Nov, Maes1991Sep}, where ${\bf x} \equiv \{x_{\alpha}\}$ is the position vector with $\alpha=1, 2, \dots , d$. Notably, however, anisotropy is not necessary for the emergence of power laws, as the algebraic decay of density correlations can also arise in isotropic systems under quite generic conditions, albeit with faster $1/r^{d+2}$ decay  \cite{Bussemaker1996Jun, Hazra2025Oct}. These observations naturally raise the question of how conservation laws and symmetries modify the fluctuation properties of such systems.
In the past, nonequilibrium absorbing-phase transitions (APTs) in mass-conserving systems—such as fixed-energy sandpile models \cite{Dickman1998May, Dickman2001Oct}—have provided a fertile ground for investigating the role of conservation laws in shaping universality classes. In particular, center-of-mass (CoM)–conserving dynamics \cite{Wilken2020Sep, Hexner2017Apr, Mukherjee2024Aug, Han2024Mar, Meerson2024Sep}, closely related to dipole-moment conservation in quantum many-body systems \cite{Morningstar2020Jun, Han2024Mar}, has attracted considerable attention \cite{Corte2008May, Corte2009Dec, Hexner2015Mar, Hexner2017Jan, Grassberger2016Oct}. Extensions to dynamics conserving higher-order moments have also been explored extensively in the literature recently \cite{Shenoy2020Feb, Morningstar2020Jun, Feldmeier2020Dec, Gromov2020Jul, Guo2022Oct, Maire2025Dec}.
While CoM conservation is known to suppress density fluctuations \cite{Hexner2017Apr, Mukherjee2024Aug, Hazra2025Feb}, whether this suppression is sufficient in the presence of anisotropy to produce {\it hyperuniformity} $-$ a novel disordered state of matter characterized by anomalously reduced long-wavelength density fluctuations \cite{Zhang1988Sep, Gabrielli2002Apr, Torquato2003Oct, Torquato2018Jun, Berthier2011Mar, Zachary2011Apr, Jack2015Feb, Vanicat2021Feb} $-$ remains an open issue. 
Indeed, the combined effects of CoM conservation and anisotropy are largely unexplored and constitute the central focus of the present work. Perhaps not surprisingly, these two competing mechanisms together are expected to shape the large-scale structure of spatio-temporal correlations in non-equilibrium systems, but the question is how: They may (i) either reinforce the slow $1/r^d$ decay of correlations characteristic of anisotropic systems or (ii) instead lead to qualitatively distinct behavior with significant suppression of fluctuations manifested as faster $1/r^{d+2}$ decay of correlations and the concomitant emergence of hyperuniformity. 

%R. L. Jack, I. R. Thompson and P. Sollich, Hyperuniformity and phase separation in biased ensembles of trajectories for diffusive systems, Phys. Rev. Lett. 114, 060601 (2015), doi:10.1103/PhysRevLett.114.060601.

%Rafael Maire, J chem Phys. for higher-moment conserving dynamics

%In this scenario, characterizing how CoM conservation modifies, or competes with, anisotropic effects is quite desirable as it could help one understand fluctuations in dynamically constrained systems, in particular those having coordinated multidirectional hopping of masses (or, particles) and exhibiting hyperuniformity. 

In this scenario, characterizing how conservation of the center of mass (CoM) modifies or competes with anisotropic effects is highly desirable, and
therefore determining which of the above two scenarios is realized—and under what conditions—is a central question addressed in this work.
Indeed, it could provide deeper insight into fluctuations in dynamically constrained systems, particularly those featuring coordinated, multidirectional hopping of masses (or particles) and exhibiting hyperuniformity.
In the past, a class of prototypical mass transport processes, with a {\it single} conserved quantity and {\it anisotropic}, reflection-symmetric hopping, have been studied extensively in the literature \cite{Garrido1990Aug, Maes1990Nov, Maes1991Sep, vanBeijeren1990Sep}. 
%%%%%%
As mentioned before, the steady-state correlation decays asymptotically as $ C(\mathbf{x}) \sim {1}/{|\mathbf{x}|^d}$ in $d$ dimensions. This $1/|\mathbf{x}|^d$ power-law scaling reflects the emergence of generic algebraic decay induced by anisotropy in the presence of only mass conservation.
%%%%%%
Here, we explore whether this scenario is modified by imposing an additional conservation law $-$ center-of-mass (CoM) conservation.
To this end, we consider minimal models to theoretically investigate the interplay between center-of-mass conservation and anisotropy. We study a broad class of conserved-mass transport processes on a $d-$dimensional hypercubic lattice with periodic boundary conditions; these systems are generalized version of Kipnis-Marchioro-Presutti (KMP) model \cite{Kipnis1982Jan, Patriarca2005, Redig2017Oct} or random average processes (RAPs) \cite{Aldous1995Jun, Coppersmith1996May, Krug2000Apr, Rajesh2000May} and have been intensively studied in the past several decades. The dynamics are reflection symmetric (thus, no net mass current) but anisotropic; notably, the anisotropy is implemented through direction-dependent hopping rates.

For simplicity, we present calculation details for two-dimensional models, where masses hop on a square lattice, and we enforce CoM conservation either along a single prescribed direction or along all spatial directions.
Later we generalize our results to higher dimensions.
Microscopically, CoM conservation is implemented through coordinated {\it multidirectional} hopping events in which two equal chunks of mass move simultaneously in opposite directions. Such pairwise updates preserve both the total mass and the center of mass [along the constrained direction(s)] and provide a minimal local mechanism for incorporating conservation of mass and its higher-order-moment. 
Our results show that CoM conservation can qualitatively alter the asymptotic behavior of correlations. 
Indeed, there are three distinct possibilities in the anisotropic systems considered in the present work.
\\\\
(i) {\it Full CoM conservation:} When CoM is conserved along {\it all} directions, the steady-state density correlation functions decay much faster, typically as
\begin{align}
 C(\mathbf{x}) \sim \frac{1}{|\mathbf{x}|^{d+2}},   
\end{align}
at large distance $|\mathbf{x}| \gg 1$, independently of whether the underlying hopping dynamics is isotropic or anisotropic. Thus, full CoM conservation significantly suppresses the $1/|\mathbf{x}|^d$ scaling characteristic of anisotropic mass-conserving systems and replaces it with a more rapidly decaying power law. Although the correlations remain algebraic, the enhanced decay implies a strong suppression of long-wavelength density fluctuations. Indeed, in Fourier space, the corresponding structure factor vanishes anomalously as the wave-number vector ${\bf q} \to {\bf 0}$ goes to zero, placing the system in the category of an extreme form of hyperuniformity $-$ referred to as ``class-I'' hyperuniformity \cite{Torquato2018Jun}, usually observed in crystals and quasi-crystals.
\\\\
(ii) {\it Partial CoM conservation:} In contrast, when CoM conservation is imposed only along a {\it single}, or some specific ({\it not} all), direction, the leading-order asymptotic behavior reverts to the slower power-law decay of correlation functions at large distance $|\mathbf{x}| \gg 1$,
\begin{align}
  C(\mathbf{x}) \sim \frac{1}{|\mathbf{x}|^d}.  
\end{align}
In this case, the partial constraint does not eliminate the dominant anisotropy-induced contribution, and the system retains the same leading power-law decay as observed in the case of mass conservation alone \cite{Garrido1990Aug, Maes1990Nov}, though with modified angular dependence.
\\\\
(iii) {\it Without CoM conservation:} Of course, when there is no CoM conservation and mass is the only conserved quantity, we recover the power-law scaling of $C(\mathbf{x}) \sim {1}/{|\mathbf{x}|^d}$ as previously found in Refs. \cite{Garrido1990Aug, Maes1990Nov, Maes1991Sep} in the context various lattice gases.

The rest of the paper is organized as follows. In Sec.~\ref{sec:models}, we define several variants of mass chipping models (MCMs) studied in this work. In Sec.~\ref{sec:calmobility}, we present the theoretical framework to calculate the transport coefficients—namely, the bulk diffusion coefficients and the Onsager coefficients (mobility tensor) for the MCMs. Using these transport coefficients, we compute the structure factor and the density–density correlation functions. In Sec.~\ref{sec:higher_d}, we obtain the asymptotic decay of the density correlation functions in higher dimensions $d > 2$. In Sec.~\ref{FDR}, we discuss a nonequilibrium version of fluctuation–dissipation relation in the context of these anisotropic systems. Finally, we summarize our results and conclude in Sec.~\ref{sec:conclusion}.

\section{Models}
\label{sec:models}

In this section, we introduce a broad class of Markov jump processes, referred to as mass chipping models (MCMs) \cite{Bondyopadhyay2012Jul, Das2016Jun, Hazra2025Oct}, which are defined on a $d$-dimensional hypercubic periodic lattice of volume $V=L^d$. These systems differ in the details of their microscopic dynamics and can be viewed as higher-dimensional generalizations of the kind of systems having fragmentation (chipping) and diffusion of masses on a lattice, such as the Kipnis–Marchioro–Presutti (KMP) model \cite{Kipnis1982Jan} and random average processes \cite{Krug2000Apr, Rajesh2000May}. Several versions of these models have been extensively studied over the past decades \cite{Aldous1995Jun, Coppersmith1996May, Patriarca2005, Redig2017Oct, Carinci2013Aug}, although predominantly in one dimension. Here, we study two (and higher) dimensional variants with a specific type of microscopic dynamics known as {\it multidirectional} hopping \cite{Hazra2025Feb, Hazra2025Oct}, in which multiple chunks of mass (or equivalently, multiple particles) simultaneously hop out of a site in different directions. We further introduce anisotropy by assigning two distinct (constant) chipping rates along the two principal lattice directions, exhibiting $C_2$ (twofold) rotational symmetry, but lacking $C_4$ (fourfold) rotational one as considered previously in Ref. \cite{Hazra2025Oct} for isotropic models. Furthermore, we incorporate an additional conservation law through center-of-mass conserving dynamics \cite{Hazra2025Oct}, wherein the two equal chunks of mass from a site are symmetrically transferred to the nearest-neighboring sites so that the local center of mass (CoM) is conserved; indeed the CoM conservation is the main focus of this study.

For simplicity, in this paper, we primarily investigate the two-dimensional case and present the details of the microscopic calculation scheme employed throughout this work. We consider a square lattice of volume $V=L^2$, where a continuous, non-negative and unbounded mass variable $m_{i,j} \ge 0$ is assigned to a site $(i,j)$. The total mass of the system, $M = \sum_{i,j} m_{i,j},$ is conserved under the stochastic dynamics, ensuring that the global density $\rho = M/L^{2}$ is fixed. In the following, we distinguish three different anisotropic variants of MCMs, classified according to the presence and nature of the center-of-mass conservation (CoMC). The first variant, referred to as {\it MCM I}, does not satisfy center-of-mass conservation. The second variant, denoted {\it CoMC IA}, preserves the center of mass along both lattice directions. The third variant, {\it CoMC IB}, conserves the center of mass only along a single lattice direction (chosen, without loss of generality, to be the 
$x-$direction on the square lattice). To compare the results obtained for multidirectional hopping, we also introduce an anisotropic model with unidirectional hopping, referred to as MCM II, where only a {\it single} ({\it not} multiple) chunk of mass hops to a randomly chosen neighboring site.

%the variant referred as \textit{MCM I}, without center-of-mass conservation; the variant \textit{CoMC IA}, with center-of-mass conservation along {\it both} lattice directions; and the variant \textit{CoMC IB}, with center-of-mass conservation along a particular (single) lattice direction (say, only along $x-$direction). 

%In these models, each site, labeled by coordinates $(i,j)$, carries a non-negative continuous mass $m_{i,j} \geq 0$. The total mass of the system, given by $M = \sum_{i,j} m_{i,j}$, is conserved. Note that, in addition to this global mass conservation, we have studied the model, in the presence of {\it anisotropy}, both with and without {\it center-of-mass conservation} (CoMC)

%\subsection{}

\paragraph{Anisotropy with a single (mass) conservation law: Model MCM I.}

\textit{Multidirectional mass transfer to all nearest neighbors and without CoM conservation:}
First, we define an anisotropic mass-conserving model that does not conserve the center of mass (CoM).
With unit rate, a site $(i,j)$ retains a fraction $(\zeta_1+\zeta_2)m_{i,j}/2$ of its mass and redistributes the remaining portion to its four nearest neighbors. The chipped-off mass is fragmented into two directional components: $\Delta m_x = \tilde{\zeta}_1 m_{i,j}/2$ along the $x$-direction and $\Delta m_y = \tilde{\zeta}_2 m_{i,j}/2$ along the $y$-direction, where $\tilde{\zeta}_k = 1 - \zeta_k$ and the chipping or the retention parameters $\zeta_k \in [0,1]$ are fixed constants ($k=1,2$). The mass $\Delta m_x$ is stochastically partitioned between the neighboring sites $(i+1,j)$ and $(i-1,j)$, which receive fractions $\xi_1 \Delta m_x$ and $(1-\xi_1)\Delta m_x$, respectively. Similarly, $\Delta m_y$ is distributed between $(i,j+1)$ and $(i,j-1)$ as $\xi_2 \Delta m_y$ and $(1-\xi_2)\Delta m_y$. The random variables $\xi_1$ and $\xi_2$ are independent and identically distributed over the unit interval $[0,1]$; generalization to arbitrary distributions is straightforward.

\paragraph{Anisotropy with full CoM conservation: Model CoMC IA.}

\textit{Center-of-mass conservation along all principal axes:}
This class of models corresponds to variants of MCM I with coordinated multidirectional hopping in which {\it center-of-mass (CoM) conservation} is strictly enforced along {\it both} the horizontal and vertical directions. With unit rate, the following two random fractions of mass $m_{i,j}$ at a site $(i,j)$ is chipped off: A chunk of mass $\Delta m_x = \xi_3 \tilde{\zeta}_1 m_{i,j}/2$ along the $x$-direction and $\Delta m_y = \xi_4 \tilde{\zeta}_2 m_{i,j}/2$ along the $y$-direction, where $\tilde{\zeta}_{\alpha} = 1-\zeta_{\alpha}$ with $\zeta_{\alpha}$ being fixed $\alpha$th direction dependent (thus anisotropic), but otherwise constant chipping parameter. The masses $\Delta m_x$ and $\Delta m_y$ are then equally split between pairs of opposing nearest neighbors to ensure a locally center-of-mass–conserving move: each of the sites $(i+1,j)$ and $(i-1,j)$ receives $\Delta m_x/2$, while each of the sites $(i,j+1)$ and $(i,j-1)$ receives $\Delta m_y/2$. The remaining mass, amounting to $(\zeta_1 + \tilde{\xi}_3 \tilde{\zeta}_1 + \zeta_2 + \tilde{\xi}_4 \tilde{\zeta}_2)m_{i,j}/2$, is retained at the departure site where $\tilde{\xi}_k = 1-\xi_k$ ($\xi_1$ and $\xi_2$ in two dimensions) are independent and identically distributed random variables, drawn from a uniform distribution $\mathcal{U}[0,1]$ in a unit interval. Throughout this paper, we denote the first and second moments as $\mu_1 = \langle \xi_k \rangle$ and $\mu_2 = \langle \xi_k^2 \rangle$, respectively.

\paragraph{Anisotropy with partial CoM conservation: Model CoMC IB.}

In this class of anisotropic models, we consider multidirectional mass transfer, but the {\it center-of-mass conservation}, unlike in CoMC IA, is now employed only along {\it a single ($x-$)direction}.
With unit rate, a site $(i,j)$ chips off directional masses $\Delta m_x = \tilde{\zeta}_1 m_{i,j}/2$ and $\Delta m_y = \tilde{\zeta}_2 m_{i,j}/2$, where $\tilde{\zeta}_k = 1-\zeta_k$ and the parameters $\zeta_k$ are fixed constants. The mass $\Delta m_x$ is redistributed symmetrically to ensure local centre-of-mass conservation along the $x$ direction: each of the sites $(i+1,j)$ and $(i-1,j)$ receives an amount $\xi_3 \Delta m_x/2$, where $\xi_3$ is a random variable. However, {\it no} center-of-mass conservation is imposed along the $y$ direction, and the mass $\Delta m_y$ is redistributed asymmetrically: a random fraction $\xi_2 \Delta m_y$ is transferred to the site $(i,j+1)$, while the remaining fraction $(1-\xi_2)\Delta m_y$ is transferred to the site $(i,j-1)$. The remaining mass $(\zeta_1/2 + \tilde{\xi}_{3} \tilde{\zeta}_{1}/2 + \zeta_2/2)$ is retained at the departure site.

\paragraph{Anisotropic, unidirectional hopping: Model MCM II.}

\begin{figure*}
    \centering
    \begin{tikzpicture}[scale=1.5, >=stealth, thick]

    % --- Panel (a) MCM I ---
    \begin{scope}[shift={(0,0)}]
        \node at (0,1.4) {\textbf{(a) MCM I}};
        \foreach \x in {-1,0,1} \foreach \y in {-1,0,1} \fill (\x,\y) circle (1.5pt);
        
        \draw[->,  ultra thick, green!60!black] (0,0) -- (0,1) node[midway, right, xshift=2pt] {$\frac{\xi_2 \tilde{\zeta}_2 m_{ij}}{2}$};
        \draw[->,  ultra thick, cyan] (0,0) -- (1,0) node[midway, below] {$\frac{\xi_1 \tilde{\zeta}_1 m_{ij}}{2}$};
        \draw[->,  ultra thick, blue!70!black] (0,0) -- (0,-1) node[midway, left] {$\frac{\tilde{\xi}_2 \tilde{\zeta}_2 m_{ij}}{2}$};
        \draw[->,  ultra thick, violet] (0,0) -- (-1,0) node[midway, above] {$\frac{\tilde{\xi}_1 \tilde{\zeta}_1 m_{ij}}{2}$};
    \end{scope}
    
    % Axis indicators between a and b
    \draw[->, black] (1.4, 0.8) -- (2.0, 0.8);
    \draw[->, black] (1.4, 0.8) -- (1.4, 1.4);
    \node at (1.9, 1.0) {$\hat{e}_x$};
    \node at (1.25, 1.4) {$\hat{e}_y$};
    \draw[dashed, gray] (1.6, 0.8) -- (1.6, -1.2);

    % --- Panel (b) CoMC IA ---
    \begin{scope}[shift={(3.4,0)}]
        \node at (0,1.4) {\textbf{(b) CoMC IA }};
        \foreach \x in {-1,0,1} \foreach \y in {-1,0,1} \fill (\x,\y) circle (1.5pt);
        
        \draw[<->,  ultra thick, green!60!black] (0,-1) -- (0,1);
        \node[green!60!black] at (0.4, 0.5) {$\frac{\xi_2 \tilde{\zeta}_2 m_{ij}}{4}$};
        \node[green!60!black] at (-0.4, -0.5) {$\frac{\xi_2 \tilde{\zeta}_2 m_{ij}}{4}$};
        
        \draw[<->, ultra thick, cyan] (-1,0) -- (1,0);
        \node[cyan] at (-0.5, 0.25) {$\frac{\xi_1 \tilde{\zeta}_1 m_{ij}}{4}$};
        \node[cyan] at (0.5, -0.25) {$\frac{\xi_1 \tilde{\zeta}_1 m_{ij}}{4}$};
    \end{scope}

    \draw[dashed, gray] (5.1, 1.5) -- (5.1, -1.2);

    % --- Panel (c) CoMC IB ---
    \begin{scope}[shift={(6.6,0)}]
        \node at (0,1.4) {\textbf{(c) CoMC IB}};
        \foreach \x in {-1,0,1} \foreach \y in {-1,0,1} \fill (\x,\y) circle (1.5pt);
        
        \draw[->, ultra thick, green!60!black] (0,0) -- (0,1) node[midway, right] {$\frac{\xi_2 \tilde{\zeta}_2 m_{ij}}{2}$};
        \draw[->,  ultra thick, blue!70!black] (0,0) -- (0,-1) node[midway, left] {$\frac{\tilde{\xi}_2 \tilde{\zeta}_2 m_{ij}}{2}$};
        \draw[<->,  ultra thick, cyan] (-1,0) -- (1,0);
        \node[cyan] at (-0.5, 0.25) {$\frac{\xi_3 \tilde{\zeta}_1 m_{ij}}{4}$};
        \node[cyan] at (0.5, -0.25) {$\frac{\xi_3 \tilde{\zeta}_1 m_{ij}}{4}$};
    \end{scope}

    % ==========================================
    % --- Panel (d) MCM II (Placed underneath) ---
    % ==========================================
    \begin{scope}[shift={(3.3,-3)}] % Centered below the top row
        \node at (0,1.4) {\textbf{(d) MCM II}};
        
        % Horizontal Moves Scenario
        \begin{scope}[shift={(-2,0)}]
            \foreach \x in {-1,0,1} \foreach \y in {-1,0,1} \fill (\x,\y) circle (1.5pt);
            
            \draw[->, ultra thick, cyan, dashed] (0,0) -- (1,0) 
              node[midway, above] {$1/2$} 
              node[midway, below,cyan] {$\tilde{\zeta}_{1}m_{ij}$};
              
            \draw[->, ultra thick, violet, dashed] (0,0) -- (-1,0) 
              node[midway, above] {$1/2$} 
              node[midway, below,violet] {$\tilde{\zeta}_{1}m_{ij}$};
        \end{scope}

        % "OR" Separator
        \node at (0,0) {\Large \textbf{OR}};

        % Vertical Moves Scenario
        \begin{scope}[shift={(2,0)}]
            \foreach \x in {-1,0,1} \foreach \y in {-1,0,1} \fill (\x,\y) circle (1.5pt);
            
            \draw[->, ultra thick, green!60!black, dashed] (0,0) -- (0,1) 
              node[midway, left] {$1/2$} 
              node[midway, right,green!60!black] {$\tilde{\zeta}_{2}m_{ij}$};
              
            \draw[->, ultra thick, blue!70!black, dashed] (0,0) -- (0,-1) 
              node[midway, left] {$1/2$} 
              node[midway, right,blue!70!black] {$\tilde{\zeta}_{2}m_{ij}$};
        \end{scope}
    \end{scope}

    \end{tikzpicture}
    
    \caption{Schematic representation of the four variants of {\it anisotropic} mass chipping models (MCMs). With unit rate, a fraction of mass $m_{i,j}$ is chipped off from a site $(i, j)$, fragmented into four chunks, and transferred with each of the fragments to one of its four nearest neighbors. The anisotropy arises from different chipping (or retention) parameters $\zeta_1$ and $\zeta_2$ along the $x$ and $y$ directions, respectively. Depending on the amount of the fragmented masses, we can define three versions of MCMs. (a) {\it MCM I:} Random {\it unequal} fractions of mass, $\xi_1 \tilde{\zeta}_1m_{i, j}/2$ and $\tilde{\xi}_1 \tilde{\zeta}_1m_{i, j}/2$, are transferred to neighbors $(i+1, j)$ and $(i-1, j)$ respectively, along the $x$-axis; a similarly fragmented mass is distributed along the $y$-axis. (b) {\it CoMC IA:} We implement a center-of-mass-conserving move in which {\it equal} amount of mass, $\xi_1\tilde{\zeta}_1m_{i, j}/4$, is transferred to both neighbors along the $x$-axis, and $\xi_2\tilde{\zeta}_2m_{i,j}/4$ is transferred to both neighbors along the $y$-axis. (c) {\it CoMC IB:} We implement a center-of-mass conserving move along the $x$-axis ($\xi_3\tilde{\zeta}_1m_{i, j}/4$ to each), but transfer of unequal chunks of mass (distributed symmetrically) along the $y$-axis. Here, $\xi_{1},\xi_{2}\in [0,1]$ are independent random variables drawn uniformly from the unit interval. (d) {\it MCM II:} A single chunk of mass is transferred either in the horizontal direction (to the left) with rate $p_{x}$ or in the vertical direction (to the right) with rate $p_{y}$. In each allowed move, the mass chips to one of its two nearest neighbors with equal probability ($1/2$). In all panels, arrows of the same color indicate equal amounts of mass being transferred, illustrating conservation of the center of mass under the specific chipping rules.}
    \label{fig:models}
\end{figure*}
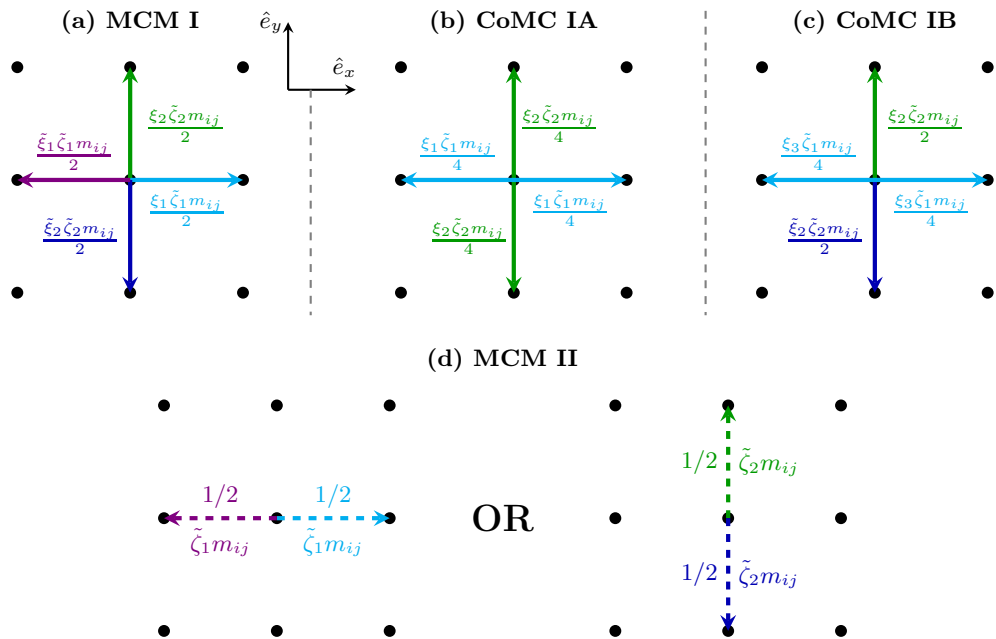

\textit{Unidirectional (a single-chunk) mass transfer:}
We also consider a class of models where only a single chunk of mass is chipped off and transferred to one of the nearest-neighbor sites. The dynamics proceed as follows. With  rate $p_{x}$, a fixed fraction $\zeta_{1}$ of the mass at a given site $(i, j)$ is retained at that site, i.e., $\zeta_{1} m_{i,j}$ remains (as shown in Fig.~\ref{fig:models}). The remaining mass, $\tilde{\zeta}_{1} m_{i,j}$, is stochastically redistributed to one of the nearest neighbors to either $(i+1, j)$ or $(i-1, j)$, and with equal probability $1/2$. With a different rate $p_{y}$ ($\ne p_x$), a fixed fraction $\zeta_{2}$ of mass at a site $(i, j)$ is retained at that site, i.e., mass $\zeta_{2} m_{i,j}$ remains at that site. The remaining mass, $\tilde{\zeta}_{2} m_{i,j}$, is stochastically redistributed to one of the nearest neighbors to either $(i, j+1)$ or $(i, j-1)$, and with equal probability $1/2$.

\section{Hydrodynamic theory and microscopic approach}
\label{sec:calmobility}

Characterizing the large-scale spatio–temporal (hydrodynamic) structure of interacting-particle systems is a central problem in statistical mechanics. On hydrodynamic scales, microscopic details become coarse-grained, and the emergent behavior is governed by conservation laws and symmetry principles. For systems with a single locally conserved quantity $-$ such as mass or particle number $-$ the macroscopic evolution is described, under diffusive scaling of space and time, within the framework of fluctuating hydrodynamics \cite{Spohn2012Dec, Bertini2015Jun}. In this description, the relevant slow variable is the coarse-grained density $\rho(\mathbf{u},\tau)$, defined through suitable (diffusive) rescaling of position $\mathbf{u} \equiv {u_\alpha}$, $\alpha=1,2,\dots,d$, and time $\tau$.
The density field evolves according to the continuity equation
\begin{align}
    \partial_{\tau} \rho({\bf u},\tau) = - \sum_{\alpha} \frac{\partial {\rm j}_{\alpha}({\bf u}, \tau)}{\partial u_\alpha},
\end{align}
which expresses local conservation of mass. The associated current ${\cal J}_{\alpha}$ can be decomposed into slow and fast contributions,
\begin{align}
\label{hyd-jd-jfl}
    {\rm j}_{\alpha}(\bf u, \tau) = {\rm j}^{(d)}({\bf u},\tau) + {\rm j}^{(\mathrm{fl})}_{\alpha}({\bf u}, \tau).
\end{align}
The first term ${\rm j}^{\rm (d)}({\bf u},\tau) = - D_{\alpha \beta} \partial_{\beta} \rho({\bf u},\tau)$ represents diffusive transport due to density gradients (Fick's law), where $D_{\alpha \beta}(\rho)$ denotes the (generally density-dependent) bulk-diffusion tensor. The second term, ${\rm j}^{(\mathrm{fl})}_{\alpha}$,  is the $\alpha-$th component of fluctuating or noise current and accounts for fluctuations arising from the underlying stochastic microscopic dynamics.
The noise current is usually taken to be Gaussian white noise, where the noise correlation can be written as 
\begin{align}
    \langle {\rm j}_\alpha^{({\rm fl})}(\mathbf{0}, 0)  {\rm j}_\beta^{({\rm fl})}(\mathbf{u}, \tau) \rangle = \frac{1}{L^d} \Gamma^{\alpha\beta}({\mathbf{u}}) \delta(\tau),
\end{align}
with $\Gamma^{\alpha \beta}$ being an element of the Onsager matrix or the mobility tensor and $L$ being a macroscopic length-scale, typically being the system size.
In the simplest case of a single conserved density (e.g., mass conservation only), the mobility reduces to $\Gamma^{\alpha\beta}(\mathbf{u})=\delta_{\alpha\beta}\delta(\mathbf{u})$, so that the noise is delta-correlated in both space and time.
By contrast, in the models considered here, the presence of an additional center-of-mass (CoM) conservation law qualitatively modifies the structure of the noise correlations. While the noise current remains delta-correlated in time, the mobility tensor acquires a spatial structure, leading to algebraic density correlations in the thermodynamic limit. Indeed, this nontrivial spatial dependence of $\Gamma^{\alpha\beta}(\mathbf{u})$ is the origin of the emergent scale-invariant spatial structures discussed in the present work.

In equilibrium, the bulk diffusion coefficient and the mobility are related by fluctuation–dissipation relations (FDRs). In nonequilibrium steady states, by contrast, such relations do not generally hold, and even when a relation between transport and fluctuation properties can be established, it may not be expressed in the standard equilibrium form. The transport coefficients in the nonequilibrium models considered here encode dynamical fluctuations intrinsic to the steady state itself. Remarkably, one can nevertheless derive a generalized nonequilibrium fluctuation–dissipation relation that connects the bulk diffusion coefficient, the mobility tensor, and the static structure factor in the small–wave-number limit [see Sec.~\eqref{FDR}].

The above hydrodynamic description can provide a universal starting point for analyzing static and dynamic correlations and large deviation properties in a diffusive system with given conservation laws and symmetries. In the present work, however, we take a direct microscopic approach to explicitly compute the transport coefficients, which in fact go into the fluctuating hydrodynamic description discussed above. 
Our goal is twofold: First, to derive the bulk-diffusion and mobility tensors starting from the underlying stochastic dynamics and, second, to characterize the emergent large-scale spatial structure in terms of these transport coefficients.
We focus on a broad class of interacting systems with {\it multidirectional} hopping, introduced in Sec.~\eqref{sec:models}, and demonstrate how their microscopic rules determine the macroscopic transport properties that enter the fluctuating hydrodynamic description.

In the following sections, we describe the microscopic dynamical framework used to exactly compute the steady-state density correlations in {\it anisotropic} mass chipping models (MCMs), both with and without center-of-mass conservation. 
First, we present the detailed calculations for a generalized variant of MCM with multidirectional hopping, referred to as MCM I, which possesses only a single conserved quantity (mass). We then discuss the main results for the other two variants of MCM, referred to as CoMC IA and CoMC IB, with constrained multidirectional hopping, where opposite pairwise movement of equal masses are incorporated and thus the CoM is also now conserved either along all directions or a particular direction, respectively.

\subsection{MCM I}

\subsubsection{Bulk-diffusion coefficient}

In this section, we study the first variant of mass-chipping models, referred to as MCM I, on a \textit{two-dimensional} hypercubic lattice of size $L \times L$. 
%%%%%%%%%%%%%%%%
%Using this model as a concrete example, we develop a general theoretical scheme for constructing a statistical mechanics framework to characterize the spatio-temporal properties of interacting particle systems.
%%%%%%%%%%%%%%
Using this model as a concrete example, we formulate a general theoretical framework for constructing a nonequilibrium statistical-mechanical description, which helps us characterize the spatiotemporal properties of interacting particle systems.
We begin by deriving the relevant transport coefficients in the model—first, the bulk diffusion coefficient, obtained from the microscopic time-evolution equation for the local mass density, and subsequently the mobility tensor, determined from the equal-time bond-current correlations.
These quantities form the central ingredients of the theoretical framework developed in the subsequent analysis.

%We illustrate the theoretical scheme to develop a general statistical mechanics framework, in the context of MCM I, to characterize the spatio-temporal properties in an interacting-particle system. We begin by deriving the transport coefficients in the model $-$ first, the bulk diffusion coefficient and then the mobility tensor $-$ from the microscopic time-evolution equation for local mass and equal-time bond-current correlations, respectively. 

To calculate the density evolution equation, we  explicitly write down the stochastic update rules for mass $m_{i,j}(t)$ at site $(i,j)$ during an infinitesimal time interval $(t, t+dt)$,
\begin{align}
\label{eq:mass_mcmI}
m_{ij}(t + \mathrm{d}t)=
\begin{cases}
\textbf{Events} & \textbf{Prob.}  \\
    m_{ij}(t) - \tilde{\zeta}_{1}\frac{m_{ij}(t)}{2} -\tilde{\zeta}_{2}\frac{m_{ij}(t)}{2}
    &  \mathrm{d}t \\
    m_{i,j}(t)+\frac{\tilde{\xi}_{1}\tilde{\zeta}_{1} m_{i+1,j}(t)}{2}
    & \mathrm{d}t\\
    m_{i,j}(t)+\frac{\xi_{1}\tilde{\zeta}_{1} m_{i-1,j}(t)}{2}
    &\mathrm{d}t \\
    m_{i,j}(t)+\frac{\tilde{\xi}_{2}\tilde{\zeta}_{2} m_{i,j+1}(t)}{2}
    & \mathrm{d}t \\
    m_{i,j}(t)+\frac{\xi_{2}\tilde{\zeta}_{2} m_{i,j-1}(t)}{2}
    &\mathrm{d}t \\
    m_{i,j}(t)
    & (1-5 \mathrm{d}t).
\end{cases}
\end{align}
Here $\xi_{1},\xi_{2}\in [0,1]$ are random variables, which, for simplicity,  are taken to be identical, independent and uniformly distributed in a unit interval. Using the above dynamical rules, the time-evolution equation for local mass can be written as
\begin{align}
\label{den-evo-mcm1}
    \frac{d}{dt}\langle m_{i,j} (t)\rangle&=
    D_{xx}(\langle m_{i+1,j}(t)\rangle+\langle m_{i-1,j}(t)\rangle-2\langle m_{i,j}(t) \rangle ) \notag \\
    &+D_{yy}(\langle m_{i,j+1}(t)\rangle+\langle m_{i,j-1}(t)\rangle-2\langle m_{i,j}(t) \rangle ),
\end{align}
where the bulk-diffusion coefficients along the horizontal and vertical directions are given by $$D_{xx} = \frac{\tilde{\zeta}_{1}}{4}; ~D_{yy} = \frac{\tilde{\zeta}_{2}}{4},$$ respectively. Here, $\tilde{\zeta}_{1}$ and $\tilde{\zeta}_{2}$ represent the chipping fractions in the $x$ and $y$ directions.
It is evident that, in general, the bulk-diffusion coefficients along two orthogonal directions, $D_{xx} \neq D_{yy}$, are different, implying that the system exhibits anisotropic diffusion. Evidently, the anisotropy arises due to the different retention or chipping parameters $\zeta_1$ and $\zeta_2$ along the two spatial directions $x$ and $y$, respectively [anisotropy can also arise due to different hopping rates along different axes as discussed in the context of a variant of MCM II with unidirectional hopping in Sec. \eqref{sec-mcmii}].

\subsubsection{Onsager coefficients or mobility tensor}

\textit{Time-integrated bond Current.} We define time-integrated bond current, along a particular axis, as the net amount of mass that has flown across a given bond in the positive direction minus the amount that has flown along the negative direction in a time interval $[0,t]$ in the steady state.
In a two-dimensional square lattice, let \( Q_{i,j}^{x}(t) \) denote the time-integrated bond current along the \( x \)-axis across a bond connecting the lattice sites \( (i,j) \) and \( (i+1,j) \). This quantity measures the cumulative net mass transferred from site \( (i,j) \) to neighboring site \( (i+1,j) \) in time \( t \) along the $x-$direction. Similarly, \( Q_{i,j}^{y}(t) \) denotes the time-integrated bond current along the \( y \)-direction across the bond connecting the neighboring sites \( (i,j) \) and \( (i,j+1) \) in time $t$.

We can write the dynamical update rules for the time-integrated current $Q^{x}_{i,j}(t)$ along the $x$-direction over the infinitesimal time interval $(t, t+dt)$ as
\begin{align}
\cQ^{x}_{i,j}(t+\mathrm{d}t) = 
\begin{cases}
    \textbf{Events} & \textbf{Prob.}\\
    \cQ^{x}_{i,j}(t)+\frac{\xi_{1}\tilde{\zeta}_{1}m_{i,j}(t)}{2} &  \mathrm{d}t \\
    \cQ^{x}_{i,j}(t)-\frac{\tilde{\xi_{1}}\tilde{\zeta}_{1}m_{i+1,j}(t)}{2} & \mathrm{d}t \\
    \cQ^{x}_{i,j}(t) & 1-2\mathrm{d}t.
\end{cases}
\end{align}
The above update rules allow us to derive the time-evolution equation for the first moment of the time-integrated bond-current $\cQ^{x}_{i,j}(t)$ along $x-$direction as follows:
\begin{align}
\label{GP1}
    \frac{d}{dt} \langle \cQ^{x}_{i,j}(t)\rangle = D_{xx}(\langle m_{i,j}(t) \rangle - \langle m_{i+1,j}(t) \rangle),
\end{align}
where the bulk-diffusion coefficient along $x-$direction is $D_{xx}=\tilde{\zeta}_{1}/4$. Similarly, from the microscopic update rule, the time evolution of the average current along the $y$-direction can be written as
\begin{equation}
\label{GP2}
    \frac{d}{dt} \langle \mathcal{Q}^{y}(x,y,t) \rangle =
      D_{yy}(\langle m_{i,j}(t) \rangle - \langle m_{i,j+1}(t) \rangle) ,
    %\simeq - D_{yy} \frac{\partial}{\partial y} m(x,y,t),
\end{equation}
where the bulk-diffusion coefficient along $y-$direction is  $D_{yy} = \tilde{\zeta}_2/4$, and the average density is defined as $\rho(x,y,t) = \langle m_{i,j}(t) \rangle$.
Under the diffusive scaling of space $(i,j) \to (u_x L, u_x L)$ and time $t \to \tau L^2$, we have the following rescaling of density and diffusive current: $m_{i,j} \to \rho(u_x, u_y)$ and diffusive current ${\cal J}_{\alpha}^{(d)}  \simeq - L^{-1} (\partial {\rm j}_{\alpha}^{(d)}/\partial u_{\alpha})$ where ${\rm j}^{\rm (d)}_{\alpha}({\bf u}, \tau) \simeq -D_{\alpha \alpha} ({\partial \rho({\bf u},\tau)}/{\partial u_{\alpha}})$ as defined in Eq. \eqref{hyd-jd-jfl}. 
Note that, for $\alpha \ne \beta$, we have  $D_{\alpha \alpha} \ne D_{\beta \beta}$ and $D_{\alpha \beta}=0$. That is, the bulk-diffusion tensor in two dimensions is simply given by
\begin{align}
    {\bf D} = 
    \begin{pmatrix}
      D_{xx} & 0 \\
      0 & D_{yy}
    \end{pmatrix},
\end{align}
where the diagonal elements are nonzero and {\it unequal} (due to anisotropy), and the off-diagonal elements are zero (thus there is no cross-diffusion).

We next derive the time-evolution equation for the equal-time two-point spatial correlation between bond currents along the $x$-direction by using the stochastic infinitesimal-time update rules given below:
\begin{widetext}
\begin{align}
\cQ^{x}_{i,j}(t+\mathrm{d}t)\cQ^{x}_{k,l}(t+\mathrm{d}t) =
\begin{cases}
\textbf{Events} & \textbf{Prob.}\\
     (\cQ^{x}_{i,j}(t)+\frac{\xi_{1}\tilde{\zeta}_{1}m_{i,j}(t)}{2})\cQ^{x}_{k,l}(t)
    & (1-\delta_{i,k} \delta_{j,l}-\delta_{i-1,k} \delta_{j,l})\mathrm{d}t ,\\
    (\cQ^{x}_{i,j}(t)-\frac{\tilde{\xi}_{1}\tilde{\zeta}_{1}m_{i+1,j}(t)}{2})\cQ^{x}_{k,l}(t)
    & (1-\delta_{ik} \delta_{jl}-\delta_{i+1,k} \delta_{j,l})\mathrm{d}t ,\\
    \cQ^{x}_{i,j}(t)(\cQ^{x}_{k,l}(t)+\frac{\xi_{1}\tilde{\zeta}_{1}m_{k,l}(t)}{2})
    &(1-\delta_{i,k} \delta_{j,l}-\delta_{i,k-1}\delta_{j,l}) \mathrm{d}t ,\\
     \cQ^{x}_{i,j}(t)(\cQ^{x}_{k,l}(t)-\frac{\tilde{\xi}_{1}\tilde{\zeta}_{1}m_{k+1,l}(t)}{2})
    &(1-\delta_{i,k} \delta_{j,l}-\delta_{i,k+1}\delta_{j,l}) \mathrm{d}t ,\\
    (\cQ^{x}_{i,j}(t)+\frac{\xi_{1}\tilde{\zeta}_{1}m_{i,j}(t)}{2})(\cQ^{x}_{k,l}(t)+\frac{\xi_{1}\tilde{\zeta}_{1}m_{i,j}(t)}{2}
    & \delta_{i,k} \delta_{j,l}\mathrm{d}t ,\\
    (\cQ^{x}_{i,j}(t)+\frac{\xi_{1}\tilde{\zeta}_{1}m_{i,j}(t)}{2})(\cQ^{x}_{k,l}(t)-\frac{\tilde{\xi}_{1}\tilde{\zeta}_{1}m_{i,j}(t)}{2})
    & \delta_{i-1,k} \delta_{j,l}\mathrm{d}t ,\\
    (\cQ^{x}_{i,j}(t)-\frac{\tilde{\xi}_{1}\tilde{\zeta}_{1}m_{i+1,j}(t)}{2})(\cQ^{x}_{k,l}(t)-\frac{\tilde{\xi}_{1}\tilde{\zeta}_{1}m_{i+1,j}(t)}{2})
    &\delta_{i,k} \delta_{j,l}\mathrm{d}t ,\\
    (\cQ^{x}_{i,j}(t)-\frac{\tilde{\xi}_{1}\tilde{\zeta}_{1}m_{i+1,j}(t)}{2})(\cQ^{x}_{k,l}(t)+\frac{\xi_{1}\tilde{\zeta}_{1}m_{i+1,j}(t)}{2})
    &\delta_{i+1,k} \delta_{j,l}\mathrm{d}t ,\\
    \cQ^{x}_{i,j}(t)\cQ^{x}_{k,l}(t)
    & 1-\Xi_{1} \mathrm{d}t.
\end{cases}
\end{align}
\end{widetext}
where $\Xi_{1}$ represents the total exit rate. 
From the above update rules, we obtain the following time-evolution equation for the equal-time bond-current correlation function:
\begin{align}
 \nonumber&\frac{d}{dt} C^{\cQ^x\cQ^x}_{r,s}(t)\\
&= D_{xx} \left[ C_{r,s}^{m\cQ^{x}} - C_{r+1,s}^{m\cQ^{x}}+C_{r,s}^{\cQ^{x}m} - C_{r-1,s}^{\cQ^{x}m} \right] +\Gamma_{r,s}^{xx}.
\end{align}
Here, we have denoted the relative position vector $(r,s)$ as $r=|k-i|$ and $s=|l-j|$, and defined the truncated correlation function $C^{\mathcal{Q}^x \mathcal{Q}^x}_{r,s}(t,t) \equiv  \langle \mathcal{Q}^{x}_{i,j}(t) \mathcal{Q}^{x}_{k,l}(t) \rangle - \langle \mathcal{Q}^{x}_{i,j}(t) \rangle \langle \mathcal{Q}^{x}_{k,l}(t) \rangle = \langle \mathcal{Q}^{x}_{i,j}(t)\mathcal{Q}^{x}_{k,l}(t) \rangle_{c}$. 
The quantity $\Gamma_{r,s}^{xx}$ represents the strength of the fluctuating (``noise'') current and can be expressed as a short-ranged function localized at the origin and nearest-neighbor sites,
\begin{equation}\label{gamma(xx)}
    \Gamma_{r,s}^{xx}=\frac{\tilde{\zeta}_{1}^{2}}{24} \langle m^{2} \rangle [4\delta(r)\delta(s)-\delta(r-1)\delta(s)-\delta(r+1)\delta(s)],
\end{equation}
where $\langle m^{2} \rangle = \int m^2 {\rm Prob.[m_i=m]} dm$ denotes the second moment of onsite mass distribution. 
Physically, the fluctuating current is defined as the difference between the instantaneous microscopic current and its slowly varying diffusive component.
It therefore represents a fast stochastic variable that captures local current fluctuations. A precise definition and a detailed discussion of its statistical properties are provided in Sec.~\eqref{FDR} and Appendix~\eqref{app:mob_cur}.

Proceeding analogously for currents along the $y$ direction, we can write the time-evolution equation for equal-time current-current correlation function, 
\begin{equation}
\label{gamma(yy)}
    \Gamma_{r,s}^{yy}=\frac{\tilde{\zeta}_{2}^{2}}{24}\langle m^{2} \rangle [4\delta(r)\delta(s)-\delta(r)\delta(s-1)-\delta(r)\delta(s+1)].
\end{equation}
The two quantities $\Gamma_{r,s}^{yy}$ and $\Gamma_{r,s}^{yy}$ in Eqs. \eqref{gamma(xx)} and \eqref{gamma(yy)}, respectively, correspond to the diagonal elements of the ``microscopic'' Onsager coefficient matrix or the mobility tensor, which are  {\it space-dependent} and characterize the strength of fluctuating or ``noise'' currents along the $x$ and $y$ directions [see Appendix~\eqref{app:mob_cur} for a precise definition of the noise current].
When integrated over space, these Onsager coefficients determine the variance of the space-integrated current and relate the bulk-diffusion coefficient to the static structure factor in the small–wave-number limit. This connection is established through a nonequilibrium fluctuation–dissipation relation (FDR), derived later in Sec.~\eqref{FDR}; see also Eq.~\eqref{eq:FDR-mobility-alpha} and Appendix~\eqref{app:mob_cur} for details.

The off-diagonal matrix elements of the ``microscopic'', {\it space-dependent} mobility tensor on the other hand can be obtained from the cross-correlations between bond currents along two orthogonal directions.
To this end, we consider the equal-time cross-correlation function $C^{\mathcal{Q}^x \mathcal{Q}^y}_{r,s}$ for orthogonal bond currents, which can be obtained from the infinitesimal-time update rules for the product of the corresponding random variables, as given below:
\begin{widetext}
\begin{align}
\cQ^{x}_{i,j}(t+\mathrm{d}t)\cQ^{y}_{k,l}(t+\mathrm{d}t) =
\begin{cases}
\textbf{Events} & \textbf{Prob.}\\
     (\cQ^{x}_{i,j}(t)+\frac{\xi_{1}\tilde{\zeta}_{1}m_{i,j}(t)}{2})\cQ^{y}_{k,l}(t)
    & (1-\delta_{i,k} \delta_{j,l}-\delta_{i,k} \delta_{j-1,l})\mathrm{d}t ,\\
    (\cQ^{x}_{i,j}(t)-\frac{\tilde{\xi}_{1}\tilde{\zeta}_{1}m_{i+1,j}(t)}{2})\cQ^{y}_{k,l}(t)
    & (1-\delta_{i+1,k} \delta_{j+1,l}-\delta_{i+1,k} \delta_{j-1,l})\mathrm{d}t ,\\
    \cQ^{x}_{i,j}(t)(\cQ^{y}_{k,l}(t)+\frac{\xi_{2}\tilde{\zeta}_{2}m_{k,l}(t)}{2})
    &(1-\delta_{i,k} \delta_{j,l}-\delta_{i,k-1}\delta_{j,l}) \mathrm{d}t ,\\
     \cQ^{x}_{i,j}(t)(\cQ^{y}_{k,l}(t)-\frac{\tilde{\xi}_{2}\tilde{\zeta}_{2}m_{k,l+1}(t)}{2})
    &(1-\delta_{i,k} \delta_{j,l+1}-\delta_{i,k-1}\delta_{j,l+1}) \mathrm{d}t ,\\
    (\cQ^{x}_{i,j}(t)+\frac{\xi_{1}\tilde{\zeta}_{1}m_{i,j}(t)}{2})(\cQ^{y}_{k,l}(t)+\frac{\xi_{2}\tilde{\zeta}_{2}m_{i,j}(t)}{2})
    & \delta_{i,k} \delta_{j,l}\mathrm{d}t ,\\
    (\cQ^{x}_{i,j}(t)+\frac{\xi_{1}\tilde{\zeta}_{1}m_{i,j}(t)}{2})(\cQ^{y}_{k,l}(t)-\frac{\tilde{\xi}_{2}\tilde{\zeta}_{2}m_{i,j}(t)}{2})
    & \delta_{i,k} \delta_{j-1,l}\mathrm{d}t ,\\
    (\cQ^{x}_{i,j}(t)-\frac{\tilde{\xi}_{1}\tilde{\zeta}_{1}m_{i+1,j}(t)}{2})(\cQ^{y}_{k,l}(t)-\frac{\tilde{\xi}_{2}\tilde{\zeta}_{2}m_{i+1,j}(t)}{2})
    &\delta_{i+1,k} \delta_{j-1,l}\mathrm{d}t ,\\
    (\cQ^{x}_{i,j}(t)-\frac{\tilde{\xi}_{1}\tilde{\zeta}_{1}m_{i+1,j}(t)}{2})(\cQ^{y}_{k,l}(t)+\frac{\xi_{2}\tilde{\zeta}_{2}m_{i+1,j}(t)}{2})
    &\delta_{i+1,k} \delta_{j,l}\mathrm{d}t ,\\
    \cQ^{x}_{i,j}(t)\cQ^{x}_{k,l}(t)
    & 1-\Xi_{2} \mathrm{d}t,
\end{cases}
\end{align}
\end{widetext}
where $\Xi_{2}$ is the total exit rate. The above dynamical rules lead to the following time-evolution equation for the equal-time correlation function for currents at two different space points:
\begin{align}
  \nonumber\frac{d}{dt} C^{\cQ^x\cQ^y}_{r,s}
&= D_{xx} \left[ C_{r,s}^{m\cQ^{y}} - C_{r+1,s}^{m\cQ^{y}} \right]\\ &
+ D_{yy} \left[ C_{r,s}^{\cQ^{x}m} - C_{r,s-1}^{\cQ^{x}m} \right]+ \Gamma_{r,s}^{xy},  
\end{align}
where  $\Gamma_{r,s}^{xy}$ is given by
\begin{align}\label{eq:gamma_xy}
    \nonumber\Gamma_{r,s}^{xy}=D_{xx}D_{yy}\langle m^{2} \rangle [\delta(r)\delta(s)-\delta(r+1)\delta(s)\\-\delta(r)\delta(s-1)+\delta(r+1)\delta(s-1)].
\end{align}
Importantly, this quantity satisfies the sum rule, 
\begin{align}
    \sum_{r,s} \Gamma_{r,s}^{xy} = 0,
\end{align}
implying that the off-diagonal (microscopic) elements do not contribute to the space-integrated (``macroscopic'' or hydrodynamic) mobility.
Consequently, the off-diagonal matrix elements of the microscopic mobility do not manifest themselves in the observable macroscopic (hydrodynamic) mobility and do not enter in a nonequilibrium  FDR relation derived later in Eq. \eqref{eq:FDR-mobility-alpha}. In this sense, although local cross-correlations between orthogonal currents are present on a microscopic scale, they do not affect the bulk transport coefficients, which can be measured on a macroscopic scale.

Now, we proceed by taking Fourier transform of Eqs.~\eqref{gamma(xx)}, \eqref{gamma(yy)}, and \eqref{eq:gamma_xy}, and we obtain the current–current 
correlation in Fourier space, which, as we see below, connects to the static structure factor. For correlations along the same direction ($\alpha = x,y$), we  write the Fourier modes of the mobility tensor in the following convenient form: The diagonal matrix elements are written as
\begin{equation}\label{gam(qx)}
    \Gamma_{q_x,q_y}^{\alpha\alpha}
    = \gamma_0^{\alpha} + \gamma_1^{\alpha}\,\lambda(q_{\alpha}),
\end{equation}
whereas for cross-correlation for currents along two orthogonal directions, or the off-diagonal matrix elements, 
($\alpha \neq \beta$) is given by
\begin{equation}\label{gam(qxy)}
    \Gamma_{q_{\alpha}, q_{\beta}}^{\alpha\beta}
    = \gamma_2^{\alpha}\gamma_2^{\beta}
    (1 - e^{i q_{\alpha}})(1 - e^{-i q_{\beta}}).
\end{equation}
In the above expressions, we have introduced the following coefficients $-$ $\gamma$'s $-$ which, in the case of MCM I, are given by 
$$\gamma_0^{\alpha}(\rho) = 2 \gamma_1^{\alpha}(\rho)
= {\frac{4}{3} D_{\alpha\alpha}^{2} \langle m^{2}\rangle}$$ 
and 
$$\gamma_2^{\alpha}(\rho) = D_{\alpha\alpha}\sqrt{\langle m^{2}\rangle},$$ 
with $\lambda(q_{\alpha}) = 2 \bigl( 1 - \cos q_{\alpha}\bigr)$, $\alpha \in \{x,y\}$.
Note that the symmetry relation 
$\Gamma^{\alpha\beta}_{q_{\alpha},q_{\beta}}=\Gamma^{\beta\alpha}_{-q_{\alpha},-q_{\beta}}$
is satisfied.
Furthermore, we can now construct a scalar quantity by summing over the elements of the Onsager coefficients matrix, or the mobility tensor, in the following way:
\begin{equation}
\label{eq:Bqgamma}
    B(\textbf{q}) = \sum_{\alpha,\beta} (1-e^{-iq_{\alpha}})(1-e^{iq_{\beta}}) \Gamma^{\alpha\beta}_{\textbf{q}} \simeq \sum_{\alpha,\beta} q_{\alpha} q_{\beta} \Gamma_{\bf q}^{\alpha \beta},
\end{equation}
where the last step has been obtained by using a small-wave-number approximation. 
Now, using Eqs. \eqref{gam(qx)} and \eqref{gam(qxy)} in Eq. \eqref{eq:Bqgamma}, we find that 
\begin{align}\label{currB:q}
     B(q_{x},q_{y})&= \langle m^{2} \rangle \Big[ \frac{\tilde{\zeta}_{1}^{2}}{24}(2\lambda(q_{x})+\lambda^{2}(q_{x}))\\ \nonumber &+\frac{\tilde{\zeta}_{2}^{2}}{24}(2\lambda(q_{y})+\lambda^{2}(q_{y}))+\frac{\tilde{\zeta}_{1}\tilde{\zeta}_{2}}{8}\lambda(q_{y})\lambda(q_{x}) \Big].
\end{align}
Next, we investigate the time evolution of equal-time spatial correlation function for local mass (density) at two different positions. We show below that the density correlation function can in fact be expressed in terms of the quantity $B_{r,s}$, where the Fourier transform of $B_{r,s}$ is provided in Eq. \eqref{currB:q}. As discussed below, $B_{r,s}$ serves as a source term in the time-evolution equation for the density correlation function.

\subsubsection{Structure factor}

In this section, we study the equal-time spatial density correlations. By using the infinitesimal-time stochastic update rules for local mass as mentioned in Eq. \eqref{eq:mass_mcmI}, we can write down the time-evolution equation for two-point density correlation function $C^{mm}_{r, s}(t, t ) = \langle m_{i, j}(t)m_{i+r, j+s}(t)\rangle -\rho^2$ as follows:
\begin{align}
\label{eq:mass}
    \frac{d}{dt}C^{mm}_{r, s}(t, t) = 2 D_{xx} \Delta_r C_{r,s}^{mm}+ 2D_{yy}\Delta_s C_{r,s}^{mm} + B_{r,s},
\end{align}
where $\Delta_r$ the discrete Laplacian operator defined as $\Delta_r C^{mm}_{r, s} = (C^{mm}_{r+1, s}+C^{mm}_{r-1, s} -2C^{mm}_{r,s})$ and the source term in the above equation can be explicitly expressed as  
\begin{align}\label{massB:r}
    B_{r,s} = \frac{\langle m^2 \rangle}{24} \bigg[ \tilde{\zeta}_1^2 (\Delta_r^2 - 2\Delta_r) + \tilde{\zeta}_2^2 (\Delta_s^2 - 2\Delta_s) \notag \\+ 3\tilde{\zeta}_1\tilde{\zeta}_2 \Delta_r \Delta_s \bigg] \delta(r)\delta(s).
\end{align}
One can readily verify that the Fourier transform of the source term $B_{r,s}$ is indeed that already given in Eq. \eqref{currB:q}. This establishes an immediate connection, via Eq. \eqref{eq:Bqgamma}, between the spatial structure and the transport coefficients, such as the bulk-diffusion coefficient and the mobility, encoding the dynamical characteristics of relaxation and fluctuation. On a physical ground, this connection between the mobility tensor [as in Eqs. \eqref{gam(qx)} and \eqref{gam(qxy)}] and the density correlations (or, equivalently, the structure factor) is quite natural: owing to mass conservation as the density and the associated current are linked through the continuity equation. As we show later, this connection can be formalized and expressed more quantitatively in terms of a nonequilibrium analogue of the fluctuation-dissipation (or, Green-Kubo) relation [see Sec. \eqref{FDR}], which is well known in the context of equilibrium systems.

Now, we use the steady-state condition $dC_{r,s}^{mm}/dt=0$ and then take Fourier transform of Eq. \eqref{eq:mass}, to obtain the following equation,
\begin{equation}
\label{Sq-MCM1}
    2S(q_{x},q_{y}) \left[ D_{xx}\lambda(q_{x})+D_{yy}\lambda(q_{y}) \right]=B(q_{x},q_{y}),
\end{equation}
for the structure factor $S(q_{x},q_{y})=\sum_{r,s}C_{r,s}^{mm}e^{i(q_xr+q_ys)}$,  which is defined as the Fourier transform of the steady-state density correlation function and the rhs of the above equation is simply the Fourier transform of the sorce term given in  Eq. \eqref{massB:r}. More specifically, we can rewrite the structure factor as the sum of three parts encoding one short-range (SR) contribution and two long-range (LR) ones,
\begin{widetext}
    \begin{align}\label{eq:sqI}
S(q_{x},q_{y}) &= \langle m^{2} \rangle \bigg[ 
 \frac{1}{2} \underbrace{  \left\{ D_{xx}\lambda(q_{x}) + D_{yy} \lambda(q_{y}) \right\} }_{SR} + \frac{2}{3} \cdot \underbrace{  \left\{ \frac{D_{xx}^{2}\lambda(q_{x}) + D_{yy}^{2}\lambda(q_{y})}{D_{xx}\lambda(q_{x}) + D_{yy}\lambda(q_{y})} \right\} }_{LR \sim 1/r^d} 
- \frac{1}{6} \cdot \underbrace{ \left\{ \frac{D_{xx}^{2} \lambda^{2}(q_{x}) + D_{yy}^{2}\lambda^{2}(q_{y})}{D_{xx}\lambda(q_{x}) + D_{yy}\lambda(q_{y})} \right\} }_{LR \sim 1/r^{d+2}} \bigg]\notag\\ &\equiv \tilde{S}_I(\textbf{q}) + \tilde{S}_{II}(\textbf{q})+\tilde{S}_{III}(\textbf{q}),
\end{align}
\end{widetext}
where $\langle m^2\rangle$ is the second moment of onsite mass and can be found from the following condition:
\begin{align}
 \langle m^2 \rangle = \left[ \frac{1}{(2\pi)^2} \int_{BZ}S(\textbf{q}) d^2\textbf{q} \right] + \rho^2.
 \end{align}
%\textcolor{red}{What is the exact expression of $\langle m^2 \rangle $ ?? Explicitly calculating $S_{II}(q)$ integral is difficult !! Maybe, we can DO it numerically and TABULATE the results? }

\begin{figure*}
\centering

\includegraphics[width=0.33\linewidth]{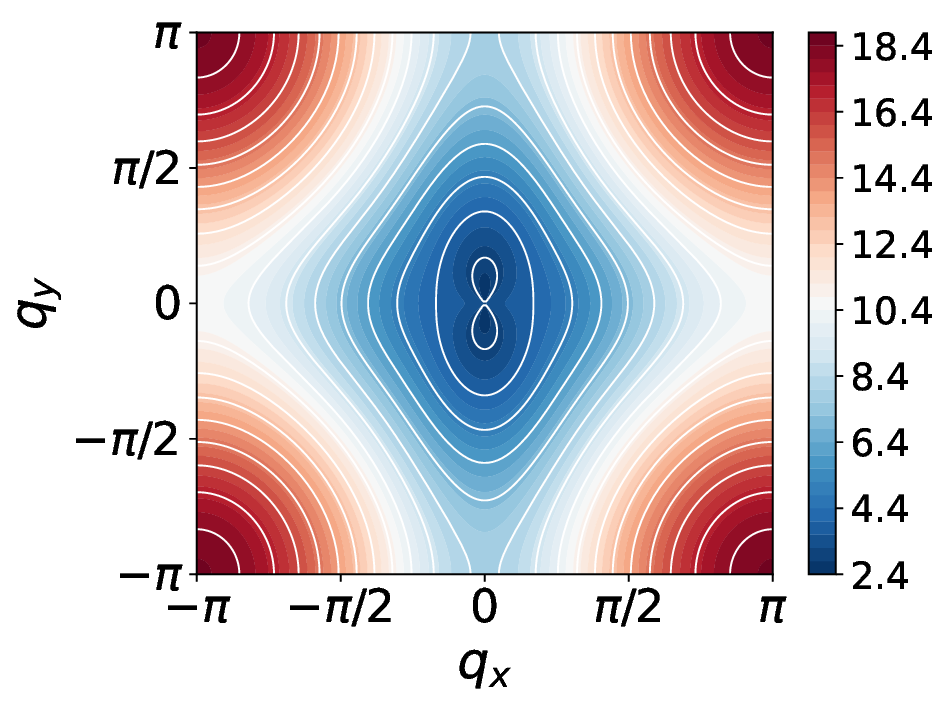}
\put(-130,110){ (a)}
\includegraphics[width=0.32\linewidth]{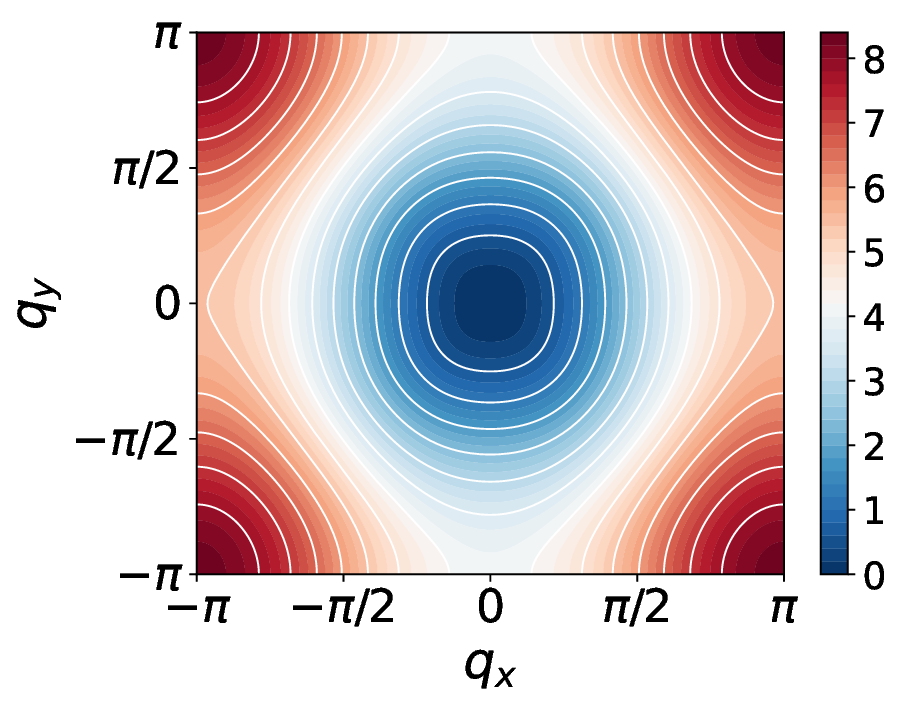}
\put(-130,110){ (b)}
\includegraphics[width=0.33\linewidth]{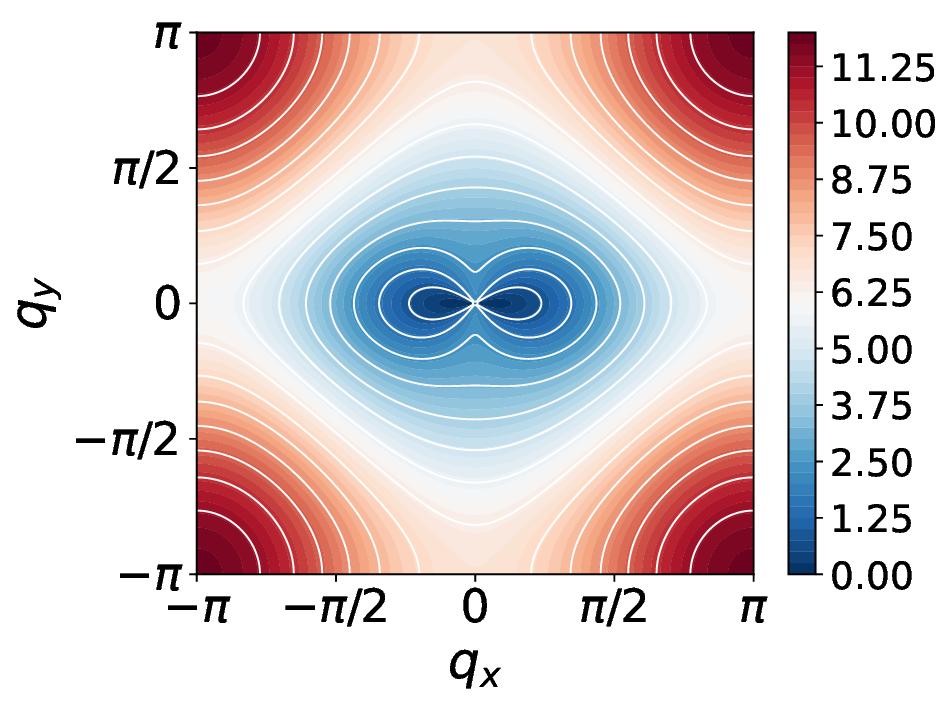}
\put(-130,110){ (c)}

\includegraphics[width=0.33\linewidth]{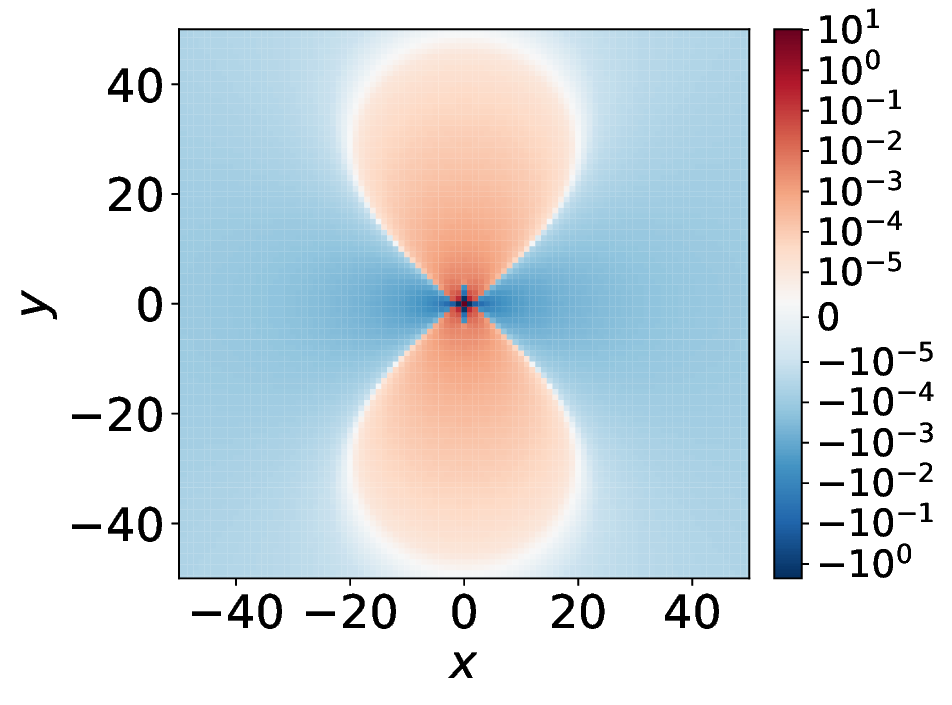}
\put(-130,110){ (d)}
\includegraphics[width=0.33\linewidth]{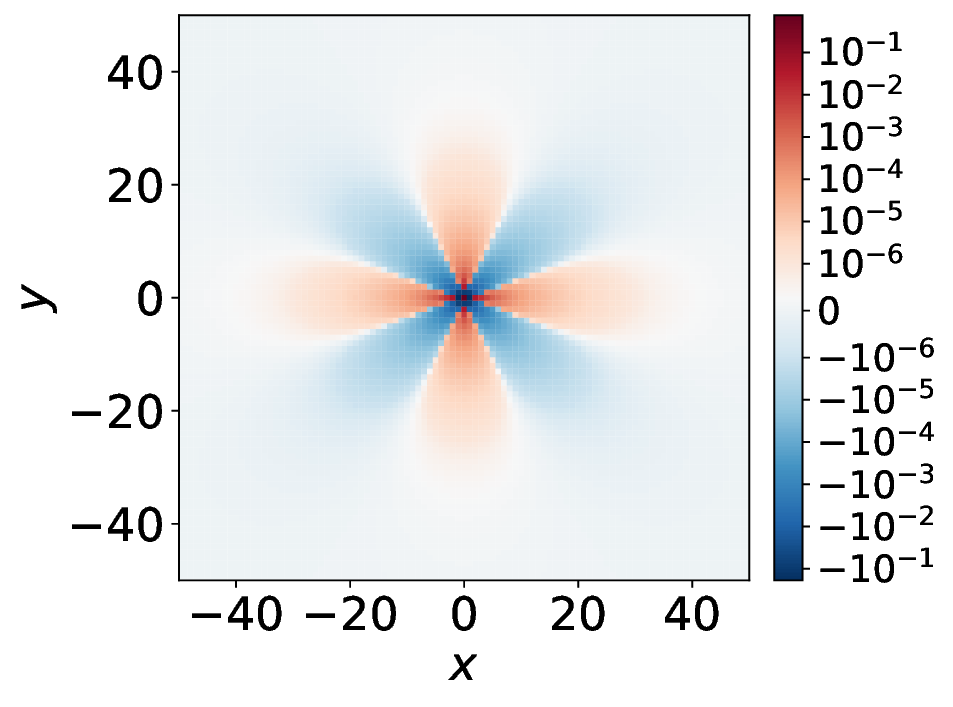}
\put(-130,110){ (e)}
\includegraphics[width=0.33\linewidth]{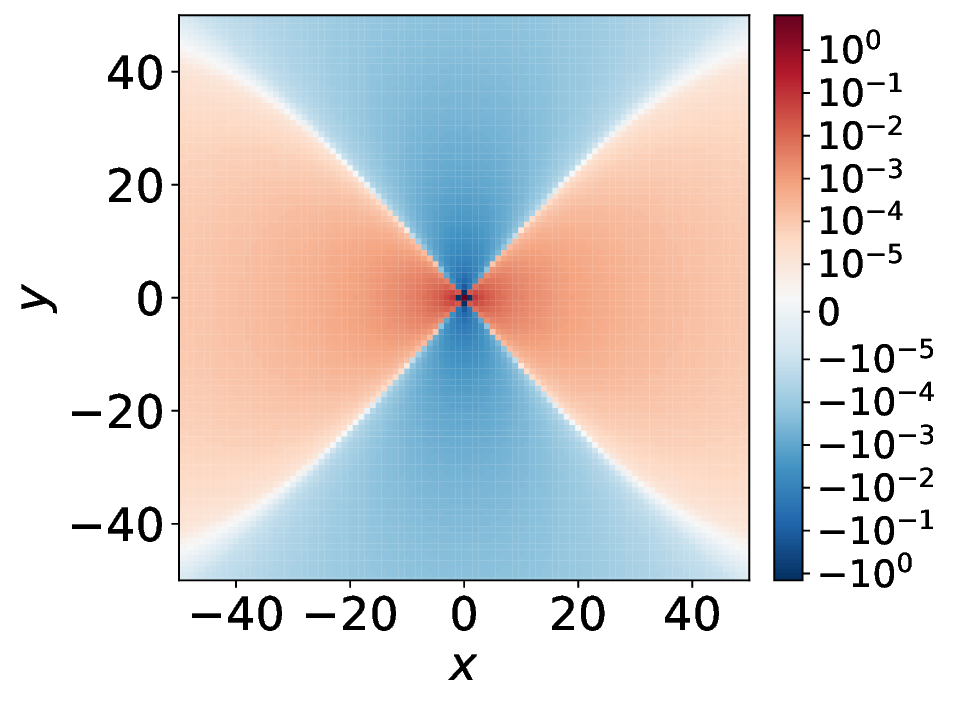}
\put(-130,110){ (f)}

\caption{
{\it Structure factor and real-space density correlation functions for several variants of mass chipping models (MCMs).}
Panels (a)–(c) show the structure factor $S(\mathbf{q})$ in the first Brillouin zone for (a) MCM-I (without center-of-mass conservation; see Eq. \eqref{eq:sqI}), (b) CoMC-IA (center-of-mass conservation along both $x-$ and $y-$ directions; see Eq. \eqref{sIA}), and (c) CoMC-IB (center-of-mass conservation along $x-$direction only; see Eq. \eqref{sB}). 
Panels (d)–(f) display the corresponding real-space heat maps of the density–density correlation function in steady state.
We take global density $\rho=4$  and chipping parameters $\zeta_1=0.2$ and $\zeta_2=0.4$ in all panels.}

\label{fig:heatmap}
\end{figure*}

\begin{figure*}
    \centering
    \includegraphics[width=0.33\linewidth]{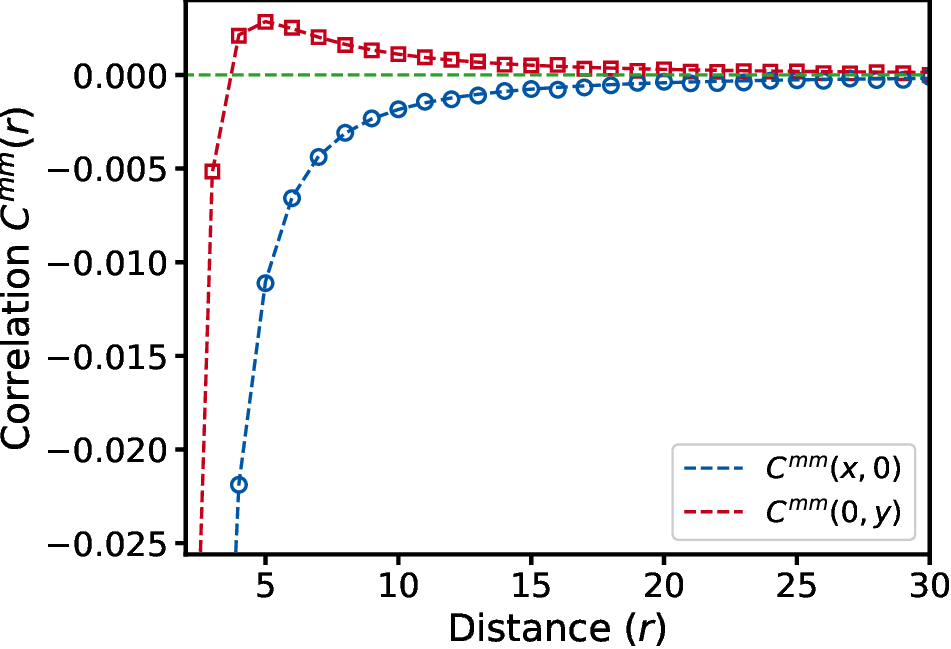}
    \put(-120,80){ (a)}
    \includegraphics[width=0.32\linewidth]{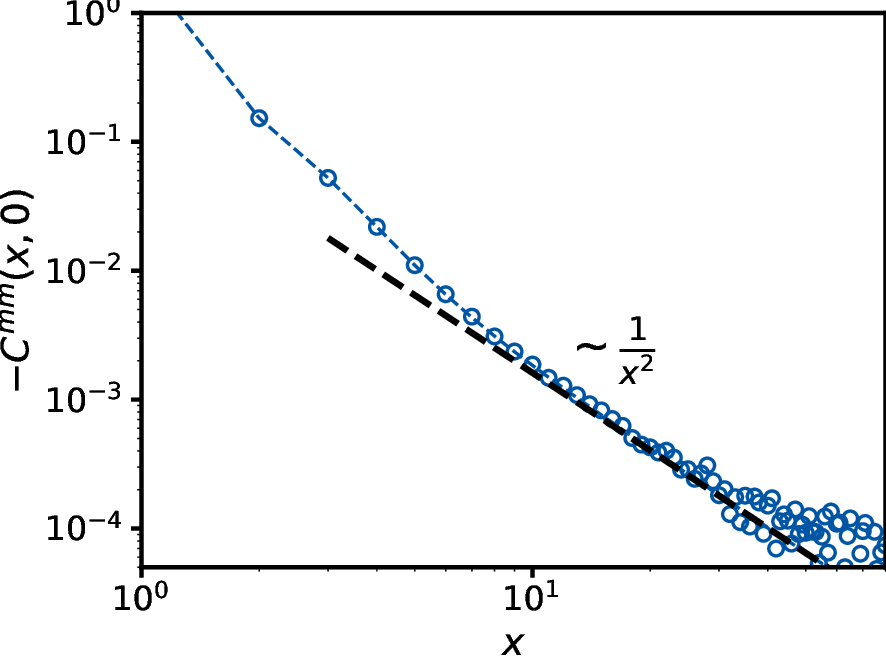}
    \put(-60,80){ (b)}
    \includegraphics[width=0.32\linewidth]{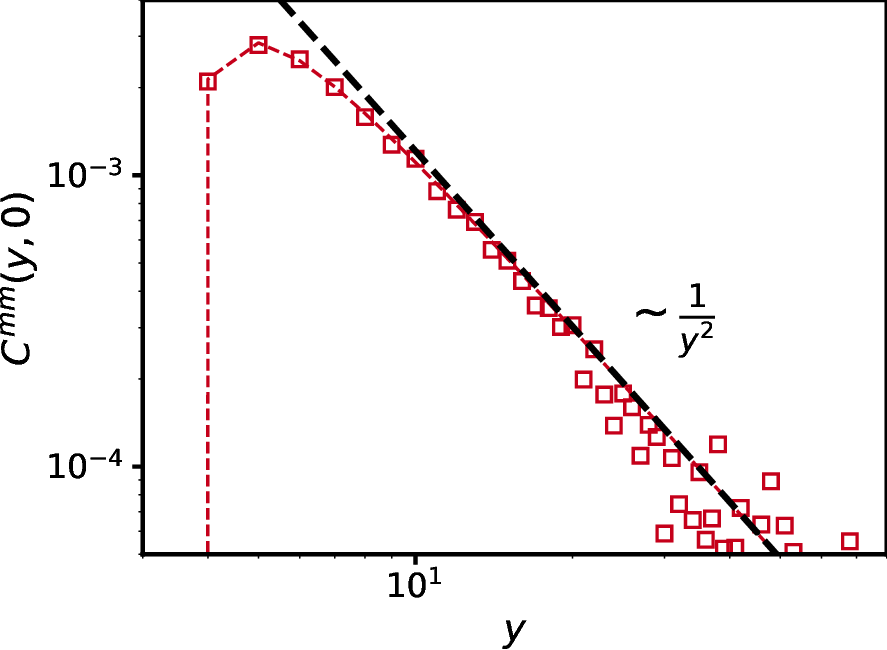}
    \put(-60,80){(c) }
    \caption{{\it Static density correlations in model variant MCM I:} (a) Spatial dependence of the correlation function $C^{mm}(\mathbf{r})$ along the principal axes. Note that the correlations exhibit sign anisotropy: they are negative along the $x$-direction (blue circles) and positive along the $y$-direction (red squares). (b) the density correlations along the $x$-axis, plotted as $-C^{mm}(x, 0)$. Similarly, density correlation  $C^{mm}(0, y)$ is plotted along the $y$-axis, $C^{mm}(0, y)$ in panel (c). In all panels, symbols represent simulation results for a system size of $300 \times 300$, while colored dashed lines correspond to the exact theoretical prediction from Eq.~\eqref{eq:cor_mcmI}. The black dashed lines in (b) and (c) indicate the asymptotic power-law decay predicted by Eqs.~\eqref{eq:mcm_x_assymp} and \eqref{eq:mcm_y_assymp}, respectively, showing excellent agreement with the data. Parameters used: $D_{xx} = 0.2$, $D_{yy} = 0.15$ and global density $\rho=4$.}
    \label{fig:crMCMI}
\end{figure*}

\subsubsection{Density correlation and its asymptotics}

In the thermodynamic limit $L \to \infty$, we can obtain the density correlation function in real space ${\bf r} \equiv \{x,y\}$ from the inverse Fourier transform of the structure factor, 
\begin{align}\label{eq:cor_mcmI}
    C^{mm}({\bf r}) =\frac{1}{(2\pi)^2} \int_{BZ} d^2{\bf q} S({\bf q}) e^{-i {\bf q}.{\bf r}} ,
\end{align}
where the integral over the wave-number has been taken on the first Brillouin zone (BZ).
For MCM I in two dimensions, we denote the correlation function as $C^{I}({\bf r} = \{x,y\})$, which can be explicitly written, after some tedious algebra, as
\begin{widetext}
\begin{align}
    C^{I}(\textbf{r})= &\langle m^{2}\rangle\sum_{\alpha=x,y} \left[ D_{\alpha \alpha}\delta(\mathbf{r})-\frac{D_{\alpha \alpha}}{2}\sum_{v \in [-1,1]}\delta(\mathbf{r}+v\hat{\mathbf{e}_{\alpha}}) \right]
    +\frac{1}{4\pi\sqrt{D_{xx}D_{yy}}(\sum_\mu x_\mu^2/D_{\mu\mu})} \sum_{\alpha=x,y}\left[ \frac{\gamma_{0}^{\alpha}}{D_{\alpha\alpha}}-\frac{2\gamma_{0}^{\alpha}x_{\alpha}^2}{D_{\alpha\alpha}^2R} \right]\notag\\
    &+\frac{3}{2\pi\sqrt{D_{xx} D_{yy}}}\frac{1}{R^2} \sum_{\alpha=x,y} \frac{\gamma_1^{\alpha} -\left(\gamma_2^{\alpha}\right)^2}{D_{\alpha\alpha}^{2}}  \left[ 1-\frac{8x_{\alpha}^2}{D_{\alpha\alpha}R}+\frac{8x_{\alpha}^4}{D_{\alpha\alpha}^2 R^2} \right];
\end{align}
\end{widetext}
where scaled distance $R=\sum_\mu x_\mu^2/D_{\mu\mu}$ for details, see Appendix~\ref{app:Green}.
The large-distance behavior of the correlation function along the principal axes can be immediately obtained from the above expression,
\begin{align}\label{eq:mcm_x_assymp}
    C^{I}_{LR}(x,0)=\frac{1}{x^2}\frac{\langle m^2 \rangle}{3\pi } \sqrt{\frac{D_{xx}}{D_{yy}}} (D_{yy}-D_{xx})+\mathcal{O}(1/x^4),
\end{align}
and
\begin{align}\label{eq:mcm_y_assymp}
    C^{I}_{LR}(0,y)=\frac{1}{y^2}\frac{\langle m^2 \rangle}{3\pi } \sqrt{\frac{D_{yy}}{D_{xx}}} (D_{xx}-D_{yy})+\mathcal{O}(1/y^4).
\end{align}
%It is now evident that, at leading order, the qualitative decay of the correlation function is provided by a power law $1/r^{2}$, with $r=\sqrt{x^2+y^2}$, along both the axes. However, the magnitude and the sign of the prefactors are different, depending on the relative values of the bulk-diffusion coefficients and directions. In other words, the correlation function can be either positive or negative, depending on the difference between $D_{xx}$ and $D_{yy}$. Furthermore, as is clear from the asymptotic forms above, in the isotropic case (i.e., when $D_{xx} = D_{yy}$), the prefactor of $1/r^2$ contribution vanishes and the power law decay is governed by a faster power law $1/r^4$ 
It is now evident that, at leading order, the correlation function exhibits a power-law decay proportional to $1/r^{2}$, where $r = \sqrt{x^{2} + y^{2}}$, along both coordinate axes. However, the magnitude and sign of the prefactors depend on the relative values of the anisotropic bulk-diffusion coefficients. In particular, the correlation function may be either positive or negative, depending on the difference between $D_{xx}$ and $D_{yy}$.
Moreover, as follows from the asymptotic expressions above, in the isotropic case (i.e., when $D_{xx} = D_{yy}$, the prefactor of the $1/r^{2}$ term simply vanishes. Therefore, the decay of the correlation function is then governed by a {\it faster}  $1/r^{4}$ power law \cite{Hazra2025Oct}, namely the sub-leading correction of the anisotropic cases as in Eqs. \eqref{eq:mcm_x_assymp} and \eqref{eq:mcm_y_assymp} in two dimensions (similarly, in $d$ dimensions).

%\section{Variants of MCM with CoM conservation}

\subsection{Model CoMC IA}

In this section, we consider a variant of the mass chipping model in which the center of mass (CoM) is conserved along {\it all spatial directions}. The analysis follows the general scheme developed in the previous section for MCM I. 
Quite remarkably, the presence of full CoM conservation leads to an anomalous suppression of fluctuations in the systems. In particular, the two-point static density correlation function exhibits algebraic decay with a universal form $ \sim 1/r^{d+2}$, even in the presence of anisotropic hopping rates. This behavior is in sharp contrast to anisotropic models (e.g., MCM I) without CoM conservation, where correlations decay as $1/r^d$
 \cite{Garrido1990Aug, Maes1990Nov, Grinstein1990Apr}. Although anisotropy does {\it not} alter the exponent in the CoM-conserving case, it remains manifest in the prefactors governing the asymptotic decay along different spatial directions.
We demonstrate the above assertions by providing below the sketch of the calculations to first obtain the mobility tensor and the anisotropic bulk-diffusion coefficients, which finally connect to the static structure factor.

{\it The mobility tensor:} We begin by determining the mobility tensor.
The stochastic update rules for the time-integrated current $Q^{x}_{i,j}(t)$, say along the $x$-direction, during an infinitesimal time interval $(t, t+dt)$ can be written as
\begin{align}
\cQ^{x}_{i,j}(t+\mathrm{d}t) = 
\begin{cases}
    \textbf{Events} & \textbf{Prob.}\\
    \cQ^{x}_{i,j}(t)+\frac{\xi_{1}\tilde{\zeta}_{1}m_{i,j}(t)}{4} &  \mathrm{d}t \\
    \cQ^{x}_{i,j}(t)-\frac{\xi_{1}\tilde{\zeta}_{1}m_{i+1,j}(t)}{4} & \mathrm{d}t \\
    \cQ^{x}_{i,j}(t) & 1-2\mathrm{d}t.
\end{cases}
\end{align}
These rules now allow us to explicitly compute the equal-time current-current correlation function at two spatial points and the microscopic mobility tensor, which is space dependent. The real-space components of the mobility tensor are short-ranged and have the following form: the diagonal ones are given by
\begin{equation}
    \Gamma_{r,s}^{xx}=\frac{\tilde{\zeta}_{1}^{2}}{16}\langle m^{2} \rangle \mu_{2}\big[2\delta(r)\delta(s)-\delta(r-1)\delta(s)-\delta(r+1)\delta(s)\big],
\end{equation}
\begin{equation}
    \Gamma_{r,s}^{yy}=\frac{\tilde{\zeta}_{2}^{2}}{16}\langle m^{2} \rangle \mu_{2}[2\delta(r)\delta(s)-\delta(r)\delta(s-1)-\delta(r)\delta(s+1)],
\end{equation}
and the off-diagonal ones by
\begin{align}
      \Gamma_{r,s}^{xy} &= \frac{\langle m^{2} \rangle \tilde{\zeta}_{1}\tilde{\zeta}_{2}}{16}\mu_{1}^{2} \big[\delta(r)\delta(s)-\delta(r+1)\delta(s)\notag\\&-\delta(r)\delta(s-1)+\delta(r+1)\delta(s-1)\big],  
      \\
      \Gamma_{r,s}^{xy} &= \Gamma_{-r,-s}^{yx}.
\end{align}
Thus, the fluctuating (noise) current correlations remain local in space, reflecting the microscopic origin of the stochastic currents.
In Fourier space, the diagonal components reduce to
\begin{equation}
    \Gamma_{q_{x},q_{y}}^{xx}=\frac{\tilde{\zeta}_{1}^{2}}{16}\langle m^{2} \rangle \mu_{2} \lambda(q_{x}),
\end{equation}
\begin{eqnarray}
    \Gamma_{q_{x},q_{y}}^{yy} &=& \frac{\tilde{\zeta}_{2}^{2}}{16}\langle m^{2} \rangle \mu_{2}\lambda(q_{y}),
 \end{eqnarray}
while the off-diagonal components are given by
\begin{eqnarray}
    \Gamma_{q_{x},q_{y}}^{xy} &=& \frac{\tilde{\zeta}_{1}\tilde{\zeta}_{2}}{16}\langle m^{2} \rangle \mu_{1}^{2}(1-e^{iq_{x}})(1-e^{-iq_{y}})
       \\
    \Gamma_{q_{x},q_{y}}^{yx} &=& \Gamma_{-q_{x},-q_{y}}^{xy} .
\end{eqnarray}
These expressions encode the anisotropy of the microscopic dynamics through the distinct coefficients associated with hopping in the $x$ and  $y$ directions.
\\\\
{\it Structure factor:}
To determine the structure factor, we next consider the infinitesimal-time stochastic update rules for the onsite mass 
$m_{i,j}(t)$. The mass evolves through local chipping and redistribution processes that conserve both total mass and the center of mass along each spatial direction; the stochastic evolution equation for mass is given by
\begin{align}
\label{eq:mass_comcIA}
m_{ij}(t + \mathrm{d}t)=
\begin{cases}
\textbf{events} & \textbf{prob.}  \\
    [m_{ij}(t) - \tilde{\zeta}_{1}\frac{m_{ij}(t)}{2} -\tilde{\zeta}_{2}\frac{m_{ij}(t)}{2}\\+\frac{\tilde{\xi_1}\tilde{\zeta_1}m_{ij}(t)}{2}+\frac{\tilde{\xi_2}\tilde{\zeta_2}m_{ij}(t)}{2}]
    &  \mathrm{d}t \\
    m_{i,j}(t)+\frac{\xi_{1}\tilde{\zeta}_{1} m_{i+1,j}(t)}{4}
    & \mathrm{d}t\\
    m_{i,j}(t)+\frac{\xi_{1}\tilde{\zeta}_{1} m_{i-1,j}(t)}{4}
    &\mathrm{d}t \\
    m_{i,j}(t)+\frac{\xi_{2}\tilde{\zeta}_{2} m_{i,j+1}(t)}{4}
    & \mathrm{d}t \\
    m_{i,j}(t)+\frac{\xi_{2}\tilde{\zeta}_{2} m_{i,j-1}(t)}{4}
    &\mathrm{d}t \\
    m_{i,j}(t)
    & (1-5 \mathrm{d}t).
\end{cases}
\end{align}
Here $\xi_{1},\xi_{2}\in [0,1]$ are random variables, which are taken to be independent, with moments $\langle \xi_1 \rangle=\langle \xi_2 \rangle=\mu_1$, $\langle \xi_1^2 \rangle=\langle \xi_2^2 \rangle=\mu_2$.
Following the procedure developed for MCM I, one obtains the static structure factor in two dimensions as given below
(for simplicity, we assumed a uniform distribution of  random numbers and therefore $\mu_{1} =  \langle \xi_k \rangle =1/2$, $\mu_{2}= \langle \xi_k^2 \rangle = 1/3$):
\begin{align}\label{sIA}
S^{IA}(q_{x},q_{y}) = \langle m^{2} \rangle \Bigg[ 
 \frac{1}{2} \underbrace{ \left\{ D_{xx}\lambda(q_{x}) + D_{yy}\lambda(q_{y}) \right\}}_{SR} \notag\\+ \frac{1}{6} \cdot \underbrace{ \left\{ \frac{D_{xx}^{2}\lambda^{2}(q_{x})+ D_{yy}^{2}\lambda^{2}(q_{y})}{D_{xx}\lambda(q_{x}) + D_{yy}\lambda(q_{y})} \right\} }_{LR \sim 1/r^{d+2}} \Bigg].
\end{align}
The first term in the above expression contributes only short-ranged (sum of spatially weighted delta-functions over the neighboring sites) correlations in real space. The second, nonanalytic term governs the long-wavelength behavior and is therefore responsible for the emergent power-law decay.
\\\\
{\it Asymptotics for density correlation function:} The large-distance asymptotic form of the density correlation function along the principal axes can be extracted from the small$-
q$ behavior of the structure factor. One obtains, for $x \gg 1$ and $y\gg 1$, the following: Along $x-$axis, the correlation function is
\begin{equation}
\label{eq:MCM1A_cx}
    C_{LR}^{IA}(x,0) \simeq \frac{\langle m^2 \rangle}{\pi}\sqrt{\frac{D_{xx}^3}{D_{yy}}}\frac{1}{x^4},
\end{equation}
and, along $y-$axis, it is
\begin{equation}
\label{eq:MCM1A_cy}
    C_{LR}^{IA}(0,y) \simeq \frac{\langle m^2 \rangle}{\pi}\sqrt{\frac{D_{yy}^3}{D_{xx}}}\frac{1}{y^4}.
\end{equation}
Several noteworthy features emerge. First, unlike in MCM I, the correlations are strictly positive along both coordinate axes, irrespective of the relative magnitudes of the anisotropic bulk-diffusion coefficients. Second, the decay is universally of the form $1/r^4$ in two dimensions (i.e., 
$1/r^{d+2}$ in general 
$d$ dimensions), in sharp contrast to the $1/r^2$ power-law behavior (or, $1/r^d$ in $d$ dimensions) observed in anisotropic MCM I without CoM conservation considered in the present work or in other systems considered previously in Refs. \cite{Garrido1990Aug, Maes1990Nov}.

Indeed, the faster decay originates directly from the additional CoM conservation law, which imposes stronger constraints on long-wavelength fluctuations in the systems. As a consequence, density fluctuations are markedly suppressed, and the systems exhibit an extreme form of hyperuniformity—specifically, ``class I'' hyperuniformity according to the classification scheme of Ref. \cite{Torquato2018Jun}.

\subsection{Model CoMC IB}

We next consider another variant of the mass chipping model with center-of-mass conservation (CoMC), referred to as CoMC IB, in which the CoM is conserved {\it only along the $x-$direction}, while the dynamics along the transverse $y-$direction remain identical to those in the original MCM I without CoM conservation. This setting allows us to understand the competing roles of anisotropy and partial CoM conservation.
As we demonstrate below, when CoM conservation is imposed only along selected directions (rather than along all principal axes), anisotropy ultimately dominates over the partial constraint (i.e., partial CoM conservation). Consequently, the static density correlation function retains the generic anisotropic long-range form,
 $$C({\bf r}) \simeq ({\rm Short-range~ contribution}) + \frac{A(\theta)}{r^d} ,$$
at large distance $r \gg 1$, where 
$A(\theta)$ is an angular function encoding the anisotropy of the microscopic hopping dynamics. Thus, in contrast to the fully CoM-conserving case discussed earlier, the decay of correlation remains $1/r^d$ power law, and hyperuniformity does not emerge.

{\it Microscopic dynamics and the transport coefficients:}
For model CoMC IB, the stochastic infinitesimal-time evolution rules for mass at a site $(i,j)$ are given by
\\\\
$m_{ij}(t + \mathrm{d}t)=$\\
\vspace{-0.55cm}
\begin{align}
\label{eq:mass_comcIB}
\begin{cases}
\textbf{event} & \textbf{prob.}  \\
    m_{ij}(t) - \tilde{\zeta}_{1}\frac{m_{ij}(t)}{2} -\tilde{\zeta}_{2}\frac{m_{ij}(t)}{2}+\frac{\tilde{\xi_3}\tilde{\zeta_1}m_{ij}(t)}{2}
    &  \mathrm{d}t \\
    m_{i,j}(t)+\frac{\xi_3\tilde{\zeta}_{1} m_{i+1,j}(t)}{4}
    & \mathrm{d}t\\
    m_{i,j}(t)+\frac{\xi_3\tilde{\zeta}_{1} m_{i-1,j}(t)}{4}
    &\mathrm{d}t \\
    m_{i,j}(t)+\frac{\tilde{\xi}_{2}\tilde{\zeta}_{2} m_{i,j+1}(t)}{2}
    & \mathrm{d}t \\
    m_{i,j}(t)+\frac{\xi_{2}\tilde{\zeta}_{2} m_{i,j-1}(t)}{2}
    &\mathrm{d}t \\
    m_{i,j}(t)
    & (1-5 \mathrm{d}t).
\end{cases}
\end{align}
The update rules reflect the fact that mass transfer along the $x-$direction preserves the center of mass, whereas transfers along $y$ follow the original non-CoM-conserving dynamics of MCM I.
From the above update rules, one can derive the time-evolution equation for local mass (density) and obtain the bulk-diffusion coefficients,  $$D_{xx}= \mu_{1} \frac{\tilde \zeta_1}{4}, ~~D_{yy} = \frac{\tilde \zeta_2}{4}.$$ 
Thus, while diffusion remains anisotropic, only the longitudinal ($x$) sector carries the imprint of CoM conservation through the moment $\mu_1$.
On the other hand, the stochastic update rules for current during an infinitesimal-time interval $(t, t+dt)$ can be written as
\begin{align}
\cQ^{x}_{i,j}(t+\mathrm{d}t) = 
\begin{cases}
    \textbf{Events} & \textbf{Prob.}\\
    \cQ^{x}_{i,j}(t)+\frac{\xi_{3}\tilde{\zeta}_{1}m_{i,j}(t)}{4} &  \mathrm{d}t \\
    \cQ^{x}_{i,j}(t)-\frac{\xi_{3}\tilde{\zeta}_{1}m_{i+1,j}(t)}{4} & \mathrm{d}t \\
    \cQ^{x}_{i,j}(t) & 1-2\mathrm{d}t.
\end{cases}
\end{align}
The corresponding stochastic update rules for the time-integrated current again allow an explicit calculation of the mobility tensor using the same scheme as for MCM I.
In real space, the matrix elements of the microscopic mobility tensor are given by the following expressions: The diagonal elements can be written as
\begin{equation}
    \Gamma_{r,s}^{xx}=\frac{\tilde{\zeta}_{1}^{2}}{16}\langle m^{2} \rangle \mu_{2}\big[2\delta(r)\delta(s)-\delta(r-1)\delta(s)-\delta(r+1)\delta(s)\big],
\end{equation}
\begin{equation}
    \Gamma_{r,s}^{yy}=\frac{\tilde{\zeta}_{2}^{2}}{24}\langle m^{2} \rangle [4\delta(r)\delta(s)-\delta(r)\delta(s-1)-\delta(r)\delta(s+1)],
\end{equation}
and the off-diagonal ones as
\begin{align}
    \Gamma_{r,s}^{xy} = \frac{\langle m^{2} \rangle \tilde{\zeta}_{1}\tilde{\zeta}_{2}}{16}\mu_{1} \big[\delta(r)\delta(s)-\delta(r+1)\delta(s)\notag\\-\delta(r)\delta(s-1)+\delta(r+1)\delta(s-1)\big],~~
 \end{align}
 and
 \begin{align}
    \Gamma_{r,s}^{yx} = \frac{\langle m^{2} \rangle \tilde{\zeta}_{1}\tilde{\zeta}_{2}}{16}\mu_{1} \big[\delta(r)\delta(s)-\delta(r-1)\delta(s)\notag\\-\delta(r)\delta(s+1)+\delta(r-1)\delta(s+1)\big],~~
 \end{align}
where $\mu_{1} = \langle \xi_{3} \rangle =  1/2$ for  uniformly distributed random variables $\xi_3 \in [0,1]$. One could readily check that the off-diagonal elements satisfy the following symmetry relation $ \Gamma_{r,s}^{yx} =   \Gamma_{-r,-s}^{xy}$.
In Fourier space, we have the diagonal elements of the microscopic mobility tensor,
\begin{equation}
    \Gamma_{q_{x},q_{y}}^{xx}=\frac{\tilde \zeta_1^2}{16} \langle m^2 \rangle \mu_2 \lambda(q_{x}),
\end{equation}
\begin{equation}
    \Gamma_{q_{x},q_{y}}^{yy}=\frac{\tilde{\zeta}_{2}^{2}}{24}\langle m^{2} \rangle [2+\lambda(q_{y})],
\end{equation} 
and the off-diagonal elements,
\begin{eqnarray}
    \Gamma_{q_{x},q_{y}}^{xy} &=& \frac{\tilde{\zeta}_{1}\tilde{\zeta}_{2}}{16}\langle m^{2} \rangle \mu_{1}(1-e^{iq_{x}})(1-e^{-iq_{y}}).
      \\
    \Gamma_{q_{x},q_{y}}^{yx} &=& \Gamma_{-q_{x},-q_{y}}^{xy} .
\end{eqnarray}
{\it Structure factor:} 
Provided the anisotropic bulk-diffusion coefficients and the mobility tensor (as calculated above), one can obtain, by using the update rules Eq. \eqref{eq:mass_comcIB}, the time-evolution equation for the equal-time spatial density correlation function, similar to that derived in Eq. \eqref{eq:mass} for MCM I. Then, imposing the steady-state condition, one can calculate the static structure factor as given below (here, we have explicitly used $\mu_{1}=1/2$ and $\mu_{2}=1/3$):
\begin{widetext}
\begin{align}
\label{sB}
S(q_{x},q_{y}) = \langle m^{2} \rangle \Bigg[ 
\frac{1}{2} \underbrace{ \left\{ D_{xx}\lambda(q_{x}) + D_{yy}\lambda(q_{y}) \right\} }_{SR} + \frac{2}{3} \cdot \underbrace{ \left\{ \frac{D_{yy}^{2}\lambda(q_{y})}{D_{xx}\lambda(q_{x}) + D_{yy}\lambda(q_{y})} \right\}}_{LR \sim 1/r^d} +\frac{1}{6} \cdot \underbrace{ \left\{ \frac{D_{xx}^{2}\lambda^{2}(q_{x}) - D_{yy}^{2}\lambda^{2}(q_{y})}{D_{xx}\lambda(q_{x}) + D_{yy}\lambda(q_{y})} \right\} }_{LR \sim 1/r^{d+2}} \Bigg].
\end{align}
\end{widetext}
As in previous cases, in the above equation, analytic (first) term contributes to only short-range correlations, whereas the nonanalytic terms (second and third ones) in the small$-q$ limit determine the power-law behavior.
From the small$-q$ expansion of Eq.~\eqref{sB}, the large-distance asymptotics of the density correlation function along the principal axes can be extracted. Along the 
$x-$axis, we have
\begin{align}
\label{eq:cx_cOmc1b}
    C_{LR}^{IB}(x,0) =  \gamma_{0}^{y} \sqrt{\frac{D_{xx}}{D_{yy}^3}} \frac{1}{4\pi x^2},
\end{align}
whereas along the $y-$axis,
\begin{align}\label{eq:cy_cOmc1b}
    C_{LR}^{IB}(0,y) =  -\gamma_{0}^{y} \frac{1}{\sqrt{D_{xx}D_{yy}}} \frac{1}{4\pi y^2}.
\end{align}
Indeed, in the present case of CoMC IB, there are several interesting features, which can be observed in the behavior of the correlation function. 
First, the correlations decay asymptotically as $1/r^2$ (in 
dimension $d=2$), consistent with the generic $1/r^d$ behavior observed in anisotropic diffusive systems such as MCM I and other models studied in Refs. \cite{Garrido1990Aug, Maes1990Nov}. Thus, partial CoM conservation is insufficient to alter the decay exponent.
Second, and more strikingly, the correlations exhibit qualitative differences in their directional asymmetry. While the decay exponent is identical along both axes as in MCM I, the sign (and magnitude) of the correlation function differs: Along the $x-$direction, the correlation remains positive, whereas along the $y-$direction, it is strictly negative. 
%This sign reversal is a direct consequence of imposing partial CoM conservation, i.e., only along the $x-$axis. The longitudinal conservation law modifies the coupling between density and current fluctuations, which in turn alters the angular structure of the long-range correlations.

Therefore, in CoMC IB, anisotropy ultimately prevails over the constraint imposed by partial CoM conservation. Although the conservation law influences amplitudes and directional characteristics, it does not sufficiently suppress long-wavelength density fluctuations to produce hyperuniformity in the systems.

%First, they decay asymptoically as a power law $\sim 1/r^2$ with $r=\sqrt{x^2+y^2}$ for $r \gg 1$, similar to MCM I and the other models studied in Refs. \cite{Garrido1990Aug, Maes1990Nov}. However, unlike in MCM I, the density correlation function now behaves qualitatively differently in the sense that the correlation along $y-$ axis, due to the CoM conservation along this particular axis, is now always negative. 

\begin{figure*}
    \centering
        \includegraphics[width=0.48\linewidth]{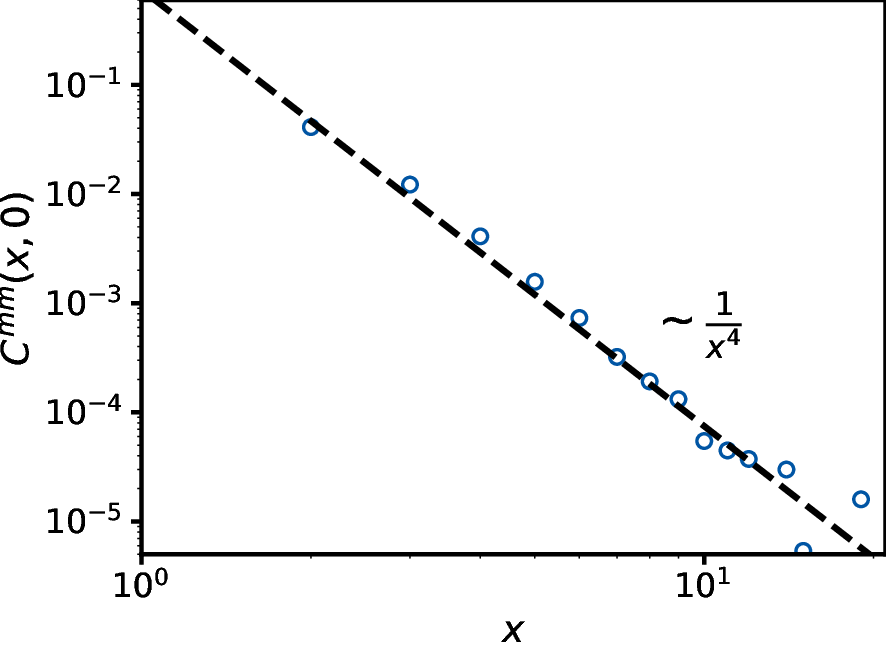}
        \put(-80,140){(a)}
        \includegraphics[width=0.48\linewidth]{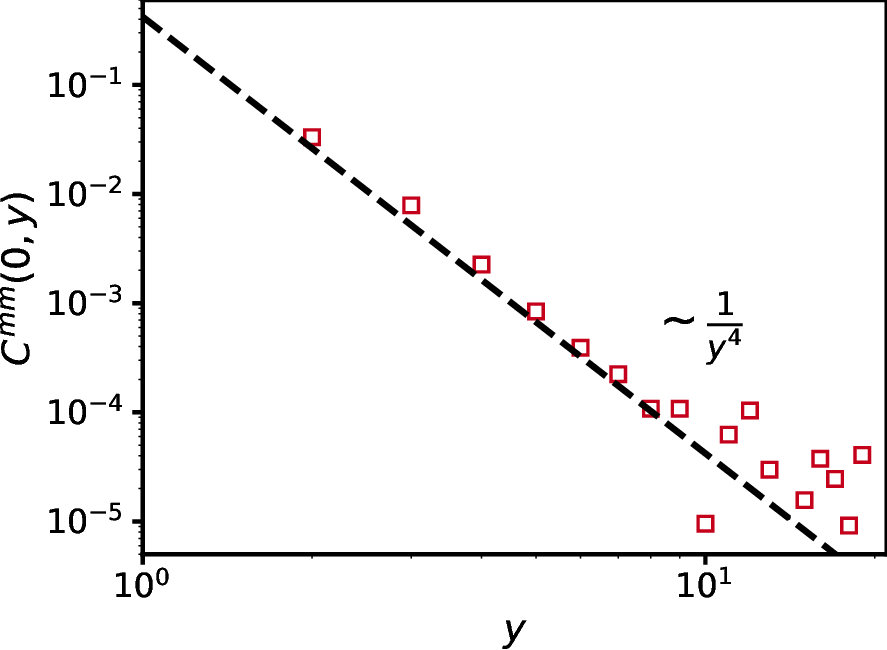}
        \put(-85,120){(b)}\\
        \includegraphics[width=0.48\linewidth]{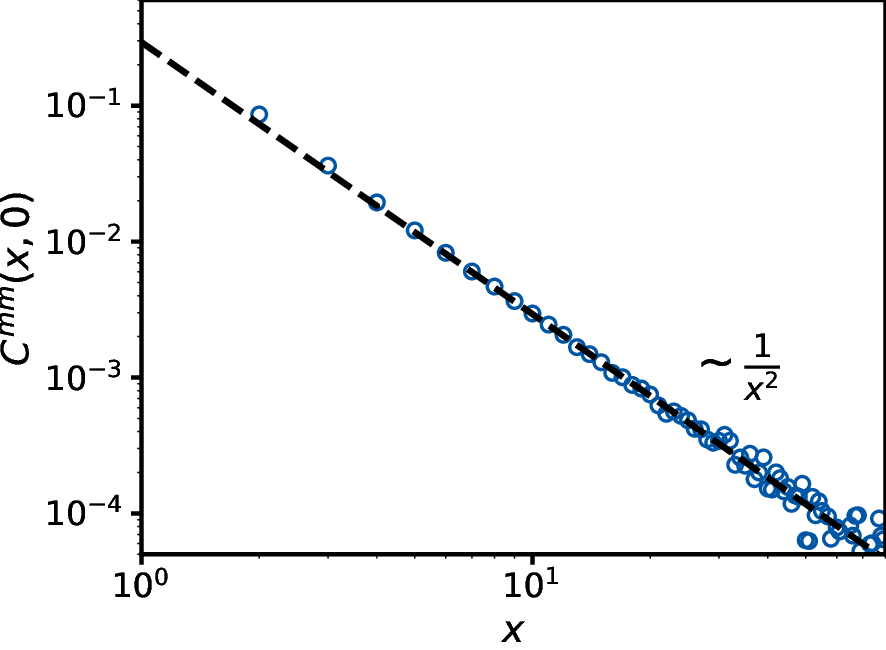}
        \put(-85,120){(c)}
        \includegraphics[width=0.48\linewidth]{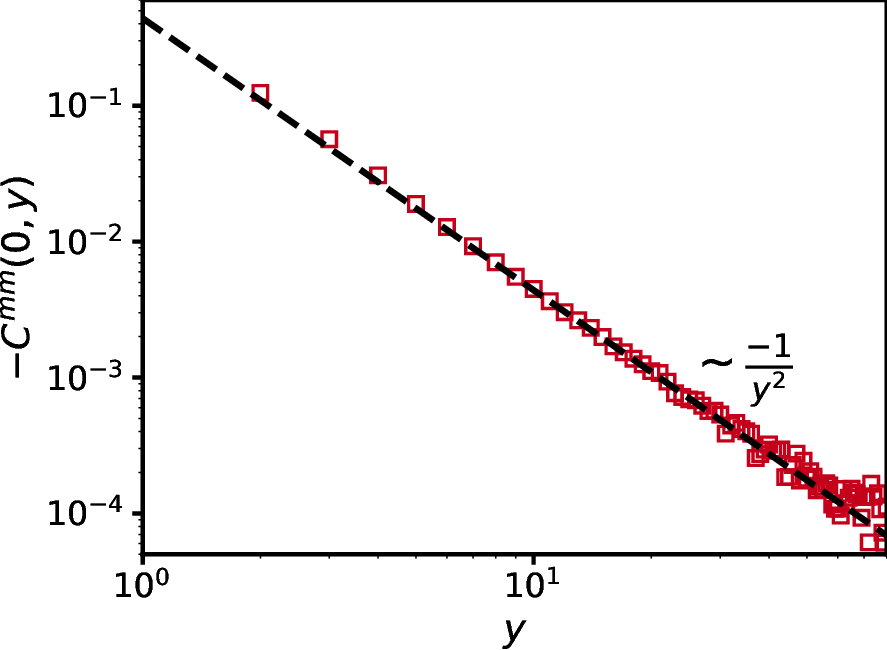}
        \put(-85,120){(d)}
    \caption{{\it CoM-conserving models $-$ CoMC IA and CoMC IB:} Density–density correlations along the $x$- and $y$-directions for two different models are shown. 
Panels (a) and (b) correspond to the model CoMC IA (center-of-mass conservation along {\it both spatial directions}), where the correlations along the $x$- and $y$-directions are plotted, respectively. The black dashed lines represent the guiding power laws $1/x^{4}$ and $1/y^{4}$, obtained from Eqs.~\eqref{eq:MCM1A_cx} and \eqref{eq:MCM1A_cy}, respectively. Panels (c) and (d) show the density correlations along the $x$- and $y$-directions, respectively, for the model CoMC IB (center-of-mass conservation only along the $x$-direction). In this case, the black dashed lines correspond to the analytical asymptotic expressions, which scale as $\sim 1/x^{2}$ and $\sim -1/y^{2}$, obtained from Eqs.~\eqref{eq:cx_cOmc1b} and \eqref{eq:cy_cOmc1b}, respectively. In all panels, the analytically obtained asymptotic forms agree well with the numerical simulation data (colored hollow circles and squares). The simulations were performed on a system of size $300 \times 300$ at global density $\rho = 4$, with parameters $\zeta_1 = 0.2$ and $\zeta_2 = 0.4$ for all panels. }
\label{fig:mcm_comc}
\end{figure*}

\subsection{Unidirectional hopping: Model MCM II}
\label{sec-mcmii}

Here, we consider anisotropic MCM with unidirectional mass-transfer rules, which would help us understand the similarities and differences between this particular version of MCM and the multidirectional ones considered in the previous sections. In model MCM II, a single chunk of mass along a particular axis, say $\alpha \in \{x,y\}$, is transferred symmetrically to the $+\alpha$ or $-\alpha$ direction with rate $p_{\alpha}$ (see detailed definition of the model in Section~\ref{sec:models}). We briefly present the calculation details and then provide the main results. 

The infinitesimal-time evolution equation for the current component, say along $x-$ direction, can be written as
\begin{align}
  \cQ^{x}_{i,j}(t+\mathrm{d}t) = 
\begin{cases}
    \textbf{Events} & \textbf{Prob.}\\
    \cQ^{x}_{i,j}(t)+\tilde{\zeta}_{1}m_{i,j}(t) & \tfrac{1}{2} p_x \mathrm{d}t \\
    \cQ^{x}_{i,j}(t)-\tilde{\zeta}_{1}m_{i+1,j}(t) & \tfrac{1}{2} p_x \mathrm{d}t \\
    \cQ^{x}_{i,j}(t) & 1-p_x\mathrm{d}t
\end{cases}
\end{align}
%The time evolution equation is;
%\begin{align}
 %   \frac{d}{dt} \langle Q^{x}_{i,j}(t)\rangle=\frac{p_1\tilde{\zeta}_{1}}{2}(\langle m_{i,j}(t) \rangle- \langle m_{i+1,j}(t) \rangle)
%\end{align}
%In general, we can write :
%\begin{equation}
 %   \frac{d}{dt} \langle Q^{\alpha}(x,y,t)\rangle=-D_{\alpha\alpha}\frac{\partial}{\partial \alpha}\rho(x,y,t) 
%\end{equation}
%where $\alpha =x,y$. 
Similar to the scheme employed for MCM I, the elements of the mobility tensor can be calculated from the equal-time current-current spatial correlation function and are given as follows: The diagonal elements can be written as
\begin{equation}
    \Gamma_{r,s}^{xx}=p_x \tilde{\zeta}_{1}^{2}\langle m^{2} \rangle\delta(r)\delta(s),
\end{equation}
\begin{equation}
    \Gamma_{r,s}^{yy}=p_y \tilde{\zeta}_{2}^{2} \langle m^{2} \rangle\delta(r)\delta(s),
\end{equation}
and, notably, unlike in MCM I, the off-diagonal elements identically vanish, i.e.,
\begin{equation}
  \Gamma_{r,s}^{xy} =  \Gamma_{r,s}^{yx} = 0.
\end{equation}
Thus, we get Fourier transform of the mobility tensor as give below: 
\begin{equation}
    \Gamma_{q_{x},q_{y}}^{xx}=p_x \tilde{\zeta}_{1}^{2}\langle m^{2} \rangle,
\end{equation}
\begin{equation}
  \Gamma_{q_{x},q_{y}}^{yy}=p_y \tilde{\zeta}_{2}^{2}\langle m^{2} \rangle  ,
\end{equation}
and
\begin{equation}
    \Gamma_{q_{x},q_{y}}^{xy}=0.
\end{equation}
The stochastic infinitesimal-time evolution equation for  mass at a site $ (i,j)$ can be written in the following way:
\begin{align}
\label{mcmii-mass-update}
m_{ij}(t + \mathrm{d}t)=
\begin{cases}
\textbf{Events} & \textbf{Prob.}\\
    m_{ij}(t) - \tilde{\zeta}_{1} m_{ij}(t) 
    & \tfrac{1}{2} p_x \mathrm{d}t + \tfrac{1}{2} p_x \mathrm{d}t, \\
    m_{i,j}(t)-\tilde{\zeta}_{2} m_{i,j}(t)
    & \tfrac{1}{2} p_y \mathrm{d}t + \tfrac{1}{2} p_y \mathrm{d}t,\\
    m_{i,j}(t)+\tilde{\zeta}_{1} m_{i+1,j}(t)
    &\tfrac{1}{2} p_x \mathrm{d}t ,\\
    m_{i,j}(t)+\tilde{\zeta}_{1} m_{i-1,j}(t)
    & \tfrac{1}{2} p_x \mathrm{d}t ,\\
    m_{i,j}(t)+\tilde{\zeta}_{2} m_{i,j-1}(t)
    & \tfrac{1}{2} p_y \mathrm{d}t ,\\
    m_{i,j}(t)+\tilde{\zeta}_{2} m_{i,j+1}(t)
    & \tfrac{1}{2} p_y \mathrm{d}t ,\\
    m_{i,j}(t)
    &1-2(p_{x}+p_{y})\mathrm{d}t.
\end{cases}
\end{align}
From the above update rules, we immediately obtain the time evolution equation for local mass (density), as given in Eq. \eqref{den-evo-mcm1},
%\begin{equation}
 %   \frac{d}{dt} \langle m_{i,j}(t) \rangle = p_x \frac{\tilde{\zeta}_{1}}{2} \frac{\partial^2}{\partial x^2} \langle m_{i,j}(t) \rangle  +p_2\frac{\tilde{\zeta }_{2}}{2}\frac{ \partial^2}{\partial y^2}\rho(x,y)
%\end{equation}
but now with the anisotropic bulk-diffusion coefficients $D_{xx}=p_x {\tilde{\zeta}_{1}}/{2}$ and $D_{yy}=p_y {\tilde{\zeta}_{2}}/{2}$.
The time-evolution of static density correlation function can be obtained from update rules in Eq. \eqref{mcmii-mass-update} and has a similar structure as that in Eq. \eqref{eq:mass}, where we have now the following expression for the source term,
\begin{align}
    B_{r,s} &=\langle m^{2} \rangle \bigg[
    p_x\tilde{\zeta}_{1}^{2}\{2\delta(r)-\delta(r+1)-\delta(r-1)\}\delta(s)\notag\\& 
    +p_y\tilde{\zeta}_{2}^{2}\delta(r)\{2\delta(s)-\delta(s+1)-\delta(s-1)\}\bigg].
\end{align}
In Fourier space, the above equation gives 
\begin{equation}
    B_{q_{x},q_{y}}=\langle m^{2} \rangle[p_{x} \tilde{\zeta}_{1}^2 \lambda(q_{x})+p_{y} \tilde{\zeta}_{2}^{2}\lambda(q_{y})],
\end{equation}
and the structure factor can then be calculated as
\begin{align}
 S(q_{x},q_{y})
 &=\frac{B_{q_{x},q_{y}}}{2D_{xx}\lambda(q_{x})+2D_{yy}\lambda(q_{y})}\\
 \nonumber &=\langle m^{2} \rangle \frac{p_{x}\tilde{\zeta}_{1}^{2}\lambda(q_{x})+p_{y}\tilde{\zeta}_{2}^{2}\lambda(q_{y})}{p_{x}\tilde{\zeta}_{1}\lambda(q_{x})+p_{y}\tilde{\zeta}_{2}\lambda(q_{y})}.
\end{align}
We conclude by highlighting a subtle aspect of the interplay between anisotropy and other microscopic details in determining the large-scale behavior of correlations.
When the chipping parameters along the orthogonal directions are {\it unequal}, $\zeta_{1} \ne \zeta_{2}$, the structure factor exhibits a nonanalyticity at origin (in Fourier space) and gives rise to $1/r^d$ power-law decay. In this regime, anisotropy in the microscopic dynamics manifests itself through direction-dependent amplitudes, but the algebraic decay remains robust.
Interestingly, the situation changes qualitatively when the chipping fractions are identical, i.e., $\zeta_{1} = \zeta_{2}$, even if the hopping rates remain anisotropic, $p_x \ne p_y$.
In this case, the steady-state structure factor reduces to a remarkably simple form,
\begin{equation} S(q_{x},q_{y})=\tilde{\zeta}\langle m^{2} \rangle,
\end{equation}
which is independent of the wave vector ${\bf q}$.
The absence of any nonanalytic contribution at small$-q$ implies that the corresponding real-space density correlation function is purely short-ranged; in fact, it reduces to a delta-function (up to a density-dependent prefactor).
This result demonstrates that anisotropy in the hopping rates ($p_x \neq p_y$) is not a {\it sufficient condition} for the emergence of power-law correlations \cite{Garrido1990Aug, vanBeijeren1990Sep}.
Rather, it is the interplay of asymmetries in the chipping parameters that generates the singular structure factor responsible for algebraic decay.
When the chipping dynamics are isotropic ($\zeta_{1} = \zeta_{2}$), the singularity is removed, and the system exhibits only short-range density correlations despite retaining anisotropic hopping rates.

%indeed, in the latter case (equal chipping fractions $\zeta_{1} = \zeta_{2}$ and unequal hopping rates $p_x \ne p_y$), one gets a short-range density correlation.

\section{Generalization to higher dimension}\label{sec:higher_d}

For the generic structure of fluctuating current in nearest-neighbor lattice models, we have the following expressions,
\begin{align}
    \Gamma^{\alpha\alpha}(\textbf{q}) = \gamma_{0}^\alpha + \gamma_1^\alpha\lambda(q_\alpha),
\end{align}
% \begin{align}
%     \Gamma^{\beta\beta}(\textbf{q}) = \gamma_{0}^\beta + \gamma_1^\beta\lambda(q_\beta),
% \end{align}
and 
\begin{align}
    \Gamma^{\alpha\beta}(\textbf{q}) =\Gamma^{\beta\alpha}(-\textbf{q})  = \gamma_{2}^\alpha\gamma^{\beta}_2(1-e^{iq_\alpha})(1-e^{-iq_\beta}).
\end{align}

\begin{table*}
\centering
\renewcommand{\arraystretch}{2.2} % Proper spacing for fractions
\setlength{\tabcolsep}{5pt}       % Optimized column spacing
\small
\begin{tabular}{l c c c c c c c c}
\toprule
\multirow{3}{*}{\textbf{Models}} & 
\multicolumn{2}{c}{\multirow{2}{*}{\textbf{Diffusivity}}} & 
\multicolumn{5}{c}{\textbf{Scaled Mobility Tensor}} &
\multirow{3}{*}{\rotatebox{90}{\textbf{Algebraic decay}}} \\
\cmidrule(lr){4-8}
 & & & \multicolumn{2}{c}{\boldmath$\frac{\gamma_{0}^{\alpha}}{(D_{\alpha\alpha})^2}$} & \multicolumn{2}{c}{\boldmath$\frac{\gamma_{1}^{\alpha}}{(D_{\alpha\alpha})^2}$} & \boldmath$\frac{\gamma_{2}^{\alpha}}{D_{\alpha\alpha}}$ & \\
\cmidrule(lr){2-3} \cmidrule(lr){4-5} \cmidrule(lr){6-7} \cmidrule(lr){8-8}
 & \boldmath$x$ ($D_{xx}$) & \boldmath$y$ ($D_{yy}$) & $x$ & $y$ & $x$ & $y$ & $x, y$ & \\
\midrule

MCM I & 
$\frac{\tilde{\zeta}_{1}}{4}$ & 
$\frac{\tilde{\zeta}_{2}}{4}$ & 
$\frac{4}{3} \langle m^{2} \rangle$ & 
$\frac{4}{3} \langle m^{2} \rangle$ & 
$\frac{2}{3} \langle m^{2} \rangle$ & 
$\frac{2}{3} \langle m^{2} \rangle$ & 
$\sqrt{\langle m^{2} \rangle}$ &
$\frac{1}{|\mathbf{x}|^d}$ \\

CoMC IA & 
$\frac{\tilde{\zeta}_{1}\mu_1}{4}$ & 
$\frac{\tilde{\zeta}_{2}\mu_1}{4}$ & 
$0$ & 
$0$ & 
$\frac{\mu_2}{\mu_1^2} \langle m^{2} \rangle$ & 
$\frac{\mu_2}{\mu_1^2} \langle m^{2} \rangle$ & 
$\sqrt{\langle m^{2} \rangle}$ &
$\frac{1}{|\mathbf{x}|^{d+2}}$ \\

CoMC IB & 
$\frac{\tilde{\zeta}_{1}\mu_1}{4}$ & 
$\frac{\tilde{\zeta}_{2}}{4}$ & 
$0$ & 
$\frac{4}{3} \langle m^{2} \rangle$ & 
$\frac{\mu_2}{\mu_1^2} \langle m^{2} \rangle$ & 
$\frac{2}{3} \langle m^{2} \rangle$ & 
$\sqrt{\langle m^{2} \rangle}$ &
$\frac{1}{|\mathbf{x}|^d}$ \\
\bottomrule
\end{tabular}
\caption{ {\it Transport coefficients in different variants of mass chipping models (MCMs).} The (anisotropic) bulk-diffusion coefficients are tabulated alongside the scaled (by $D_{\alpha\alpha}$) elements of the mobility tensor (or, the Onsager matrix), along with the asymptotic algebraic decay of the corresponding density correlation functions, in each of the variants of mass-chipping models (MCMs) studied in the present work. Here, we have denoted $\mu_1= \langle \xi \rangle$ and $\mu_2= \langle \xi^2 \rangle$ where the averages are taken, for simplicity, over the uniform distribution function $\phi(\xi)=1$ and $\xi \in [0,1]$. }
\label{tab:mobility_tensor}
\end{table*}
In general, we have an expression for the structure factor in $d-$dimensional models as given below:
\begin{align}\label{eq:Sq}
     S(\textbf{q}) 
    =\nonumber \frac{1}{2\sum_\alpha D_{\alpha\alpha}\lambda(q_\alpha)} \Bigg( \sum_{\alpha}\gamma_{0}^{\alpha}\lambda(q_{\alpha})+\Big[\sum_{\alpha}\gamma_{2}^{\alpha}\lambda(q_{\alpha})\Big]^2\\ +\sum_{\alpha}(\gamma_{1}^{\alpha}-(\gamma_{2}^{\alpha})^{2})\lambda^{2}(q_{\alpha}) \Bigg).
\end{align}
Then, the nonanalytic terms in the structure factor will give rise to power laws in the density correlation function. In the above equation, the first term in the structure factor is of order ${\cal O}(1)$ and both the second and third terms contribute to order ${\cal O}(q^2)$. As one can see, all three terms here can be nonanalytic and, in general, give rise to power-law correlations. While the ${\cal O}(1)$ first term usually gives rise to $1/r^d$ power-law decay of the correlation function, both the second and third terms, which are of the same order, in that case, give a faster $1/r^{d+2}$ power-law decay. 
In the models studied in this paper, we see that, in multi-directional hopping whenever the strength $\Gamma^{xy}$ of correlations between currents in orthogonal directions is nonzero, there is a relation between $\gamma_{2}^{\alpha}$ and $D_{\alpha \alpha}$,
\begin{equation}
    \gamma_{2}^{\alpha} = D_{\alpha\alpha}\sqrt{\langle m^{2} \rangle}.
\end{equation}
The above relation implies that the numerator of the second term in \eqref{eq:Sq} is then can be written as square of the denominator term, thus leading to the cancellation of the denominator. This results in an analytic term of order ${\cal O}(q^2)$ and consequently a short-range density correlation. In other words,, the structure factor in that case can be written as
\begin{widetext}
\begin{eqnarray}\label{f:sq}
     S(\textbf{q})
    &=& \underbrace{ \left[ \frac{\sum_{\alpha}\gamma_{0}^{\alpha}\lambda(q_{\alpha})}{{2\sum_\alpha D_{\alpha\alpha}\lambda(q_\alpha)}} \right] }_{LR \sim 1/r^d} + \underbrace{ \left[ \frac{\langle m^{2}\rangle}{2}\sum_{\alpha}D_{\alpha \alpha}\lambda(q_{\alpha}) \right] }_{SR} + \underbrace{ \left[ \frac{\sum_{\alpha}(\gamma_{1}^{\alpha}-(\gamma_{2}^{\alpha})^{2}) \lambda^{2}(q_{\alpha})}{{2\sum_\alpha D_{\alpha\alpha}\lambda(q_\alpha)}} \right] }_{LR \sim 1/r^{d+2}} ~~~~~
    \\
    &\sim & {\cal O}(1) + {\cal O}(q^2)  + {\cal O}(q^2),
\end{eqnarray}
\end{widetext}
where the first and third contributions in the square brackets are nonanlytic and lead to power-law decays ($1/r^d$ and $1/r^{d+2}$, respectively) and the second term in the square bracket gives rise to a short-ranged correlation.

The correlation function can be obtained by performing inverse Fourier transform of the structure factor Eq.\eqref{f:sq} and we write it as
\begin{widetext}
\begin{equation}
\label{corr}
    C(\mathbf{r})=\underbrace{C_{A1}(\mathbf{r})}_{\text{LR}\sim 1/r^d} + \langle m^{2}\rangle \sum_{\alpha=1}^{d} \underbrace{ \left[ D_{\alpha \alpha}\delta(\mathbf{r})-\frac{D_{\alpha \alpha}}{2}\sum_{v \in [-1,1]}\delta(\mathbf{r}+v\hat{\mathbf{e}_{\alpha}}) \right] }_{SR}  + \underbrace{C_{A3}(\mathbf{r})}_{LR \sim 1/r^{d+2}}.
\end{equation}
\end{widetext}
After doing some straightforward algebraic manipulations (see Appendix~\ref{app:Green} for details), the long-range (LR) contributions can be expressed explicitly as a function of position and all parameters of the models as given below:
\begin{align}
    C_{A1}(\mathbf{r}) = \frac{\Gamma(d/2)}{4 \pi^{d/2} \sqrt{\prod D_{\mu\mu}}} \frac{1}{R^d} \sum_{\alpha=1}^d \frac{\gamma_0^\alpha}{D_{\alpha\alpha}} \left[ 1 - d \left(\frac{X_\alpha}{R}\right)^2 \right],
\end{align}
and
%\begin{widetext}
    \begin{align}
    C_{A3}(\mathbf{r}) = \frac{3 \Gamma(d/2 + 2)}{(d+2)\pi^{d/2} \sqrt{\prod D_{\mu\mu}}} \frac{1}{R^{d+2}} \sum_{\alpha=1}^d \frac{\gamma_1^\alpha - (\gamma_2^\alpha)^2}{D_{\alpha\alpha}^2}\times \notag\\ \Bigg[ 1- 2(d+2)\left(\frac{X_\alpha}{R}\right)^2 + \frac{(d+2)(d+4)}{3}\left(\frac{X_\alpha}{R}\right)^4 \Bigg].
\end{align}
%\end{widetext}
In the above expressions, we have used a rescaled position vector $\mathbf{R} \equiv \{ X_\alpha \}$ and the magnitude $R = \sqrt{\sum_\alpha X_\alpha^2}$, where the new coordinate
$X_\alpha = x_\alpha/\sqrt{D_{\alpha\alpha}}$ is obtained by scaling the original position coordinates.
%\textcolor{red}{Furthermore, in the models studied here, whenever the strength of fluctuating currents in the orthogonal directions is zero (i.e., $\gamma_{2}^{x}=\gamma_{2}^{y}=0$)} and the second term in Eq. \eqref{eq:Sq} is absent,
%\begin{equation}
 %    C(\textbf{r})   =\int d\textbf{q}\Bigg[\frac{\sum_{\alpha}\gamma_{0}^{\alpha}\lambda(q_{\alpha})}{{2\sum_\alpha D_{\alpha\alpha}\lambda(q_\alpha)}}+\frac{\sum_{\alpha}(\gamma_{1}^{\alpha}-(\gamma_{2}^{\alpha})^{2})\lambda^{2}(q_{\alpha})}{{2\sum_\alpha D_{\alpha\alpha}\lambda(q_\alpha)}} \Bigg]e^{i\textbf{q}.\textbf{r}},
%\end{equation}
%\textcolor{red}{XXXXX this sentence seems incomplete ...}

\section{A nonequilibrium fluctuation-dissipation relation} 
\label{FDR}

Although detailed balance is violated at the microscopic level, the system exhibits emergent large-scale structure that somewhat resemble those in equilibrium. In particular, one can identify a nonequilibrium version of the familiar equilibrium fluctuation-dissipation relations that connect transport coefficients to static correlations in the hydrodynamic limit.
To establish the fluctuation relations, we first introduce the macroscopic (hydrodynamic) mobility, which can be expressed in terms of
the structure factor $S(\mathbf{q} \to 0)$ in the small wave-number limit. 
In two (and higher) dimensions, this limit is generally {\it path dependent} (or, in other words, the ``compressibility'' is {\it shape dependent}). Therefore, one may approach either by taking $q_\alpha \to 0$ first, followed by $q_\beta \to 0$, or vice versa. Accordingly, we can define the following quantities, depending on the order of the limits taken in two orthogonal directions $\alpha \neq \beta$:
\begin{equation}
\label{S0-1}
    S^{(0)}_{\alpha} \equiv \lim_{q_\alpha \to 0} \Big[\lim_{q_\beta \to 0} S(\textbf{q}) \Big] = \lim_{q_\alpha \to 0} S(q_{\alpha}, q_{\beta}=0) = \frac{\gamma_{0}^{\alpha}}{2D_{\alpha \alpha}} ,
\end{equation}
and
\begin{equation}
\label{S0-2}
     S^{(0)}_{\beta} \equiv \lim_{q_\beta \to 0}\Big[\lim_{q_\alpha \to 0}S(\textbf{q})\Big] = \lim_{q_\beta \to 0} S(q_{\alpha}=0, q_{\beta}) =\frac{\gamma_{0}^{\beta}}{2D_{\beta \beta}} .
\end{equation}
Interestingly, in certain cases even with anisotropic hopping, e.g., in CoMC IA and MCM II with unidirectional chipping of a single chunk of mass, one can have a relation,
\begin{align}\label{eq:FDR}
    S^{(0)}_{\alpha}  =  S^{(0)}_{\beta},
\end{align}
implying that the ratios in Eqs. \eqref{S0-1} and \eqref{S0-2} are equal, and consequently {\it no} nonanalytic contribution arises at order ${\cal O}(q^0)$ in the structure factor (as in Eq. \eqref{eq:Sq}). Such behavior has been reported previously in related nonequilibrium models \cite{vanBeijeren1990Sep}. As a result, an $1/r^d$ algebraic decay of density correlation is not possible even in the presence of anisotropy. In other words, to observe a $1/r^d$ decay, one requires the relation Eq.\eqref{eq:FDR} to be violated.
Notably, the relation Eq. \eqref{eq:FDR} is satisfied for isotropic hopping rates \cite{Hazra2025Oct}.

We now proceed to derive a nonequilibrium ``fluctuation-dissipation'' relation, somewhat analogous to the equilibrium fluctuation-dissipation relation (or, equivalently, the Green-Kubo relation), which provides a fundamental connection between current fluctuation (or, the mobility), density relaxation (or, the bulk-diffusion coefficient) and the structure factor in anisotropic systems. To this end, we first define the following quantities in the small-${\bf q}$ limit $-$ a ``macroscopic'' (coarse-grained or hydrodynamic) mobility tensor, whose components are provided below: The diagonal one is given by
\begin{align}
    \chi^{\alpha \alpha} = \frac{1}{2} \lim_{q_\alpha \to 0} \Gamma^{\alpha \alpha}(q_{\alpha}),
\end{align}
and the off-diagonal one ($\alpha \ne \beta$),
\begin{align}
\chi^{\alpha\beta} \equiv \frac{1}{2} \lim_{q_\beta \to 0} \lim_{q_\alpha \to 0} \Gamma^{\alpha\beta}(\mathbf{q}) = \lim_{q_\beta \to 0} \left[ \lim_{q_\alpha \to 0}  D_{\alpha\beta}S(\mathbf{q}) \right]\notag\\ = \lim_{q_\alpha \to 0} \left[ \lim_{q_\beta \to 0}  D_{\alpha\beta}S(\mathbf{q}) \right],
\end{align}
which vanishes as we have $D_{\alpha\beta} = D_{\beta\alpha} = 0$ for $\alpha \ne \beta$ (i.e., there is no cross-diffusion) and 
\begin{align}
D_{\alpha\beta} S^{(0)}_{\beta} =  D_{\alpha \beta} S^{(0)}_{\alpha} = 0.
\end{align}
This is consistent with the left-hand side of the above equation, where the spatial sum of the fluctuating current strengths along two orthogonal directions vanishes. Importantly, in the class of models considered here, the result does not depend on the order of the limits, $q_\alpha \to 0$ and $q_\beta \to 0$, implying the absence of cross-coupling in the bulk-diffusion tensor. 
Furthermore, the space-integrated diagonal elements of the space-dependent mobility tensor, along a particular direction $\alpha$ can be related to the variance of the total time-integrated bond current, summed over the entire system and scaled by the spacetime volume in the thermodynamic limit. that is, we have 
\begin{align}
\label{curr-fluc-mob}
    \chi^{\alpha\alpha} &=\lim_{T\to \infty, V\to \infty}\frac{1}{2d TL^d}\left\langle\left[\sum_{\textbf{r}}\cQ_{\alpha}(\textbf{r}, T)\right]^2\right\rangle \notag\\&= \frac{1}{2d}\sum_{\textbf{r}}\Gamma^{\alpha\alpha}_{\textbf{r}},
\end{align}
where $\mathcal{Q}_{\alpha}(\mathbf{r}, T)$ denotes the time-integrated bond current at position ${\bf r}$; see Appendix \ref{app:mob_cur} for the details.
Thus, the hydrodynamic mobility is directly proportional to the variance of the total space-time-integrated current, scaled by the space-time volume.
In anisotropic case, the diagonal components $\chi^{\alpha \alpha}$ of the macroscopic mobility tensor in two orthogonal directions are different.

Finally we obtain the FDR by noting that the macroscopic mobility in the above equation can be related to the structure factor by taking small$-{\bf q}$ limits along the the orthogonal axes $\alpha \ne \beta$ in the two following ways:
\begin{align}
    \chi^{\alpha \alpha} &= \frac{1}{2} \lim_{q_\alpha \to 0} \Gamma^{\alpha \alpha}(q_{\alpha}) = \lim_{q_\alpha \to 0}  D_{\alpha\alpha} S(q_{\alpha}, \{q_{\beta}=0\}) = \frac{\gamma_{0}^{\alpha}}{2}, \label{eq:FDR-mobility-alpha} \\
    \chi^{\beta \beta} &= \frac{1}{2} \lim_{q_\beta \to 0} \Gamma^{\beta \beta}(q_{\beta}) =  \lim_{q_\beta \to 0}  D_{\beta \beta} S(q_{\alpha}=0, q_{\beta}) = \frac{\gamma_{0}^{\beta}}{2}; \label{eq:FDR-mobility-beta}
\end{align}
i.e., in the first case, we first take the limit $\lim_{q_\beta \to 0}$ and then the limit $\lim_{q_\alpha \to 0}$, and, in the second case, the vice versa. 
Equivalently, we can cast the above equations analogous to the equilibrium FDR, albeit now in a compact matrix form as given below:
\begin{align}
    {\bf \chi} = {\bf D} {\bf S}^{(0)} ,
\end{align}
where we denote the matrices,
\begin{align}
 {\bf \chi} =
  \begin{pmatrix}
      \chi^{xx} & 0 \\
      0 & \chi^{yy}
    \end{pmatrix} ; ~
    {\bf D} = \begin{pmatrix} 
      D_{xx} & 0 \\
      0 & D_{yy}
    \end{pmatrix} ;~
    {\bf S}^{(0)} =
    \begin{pmatrix}
      S^{(0)}_{x} & 0 \\
      0 & S^{(0)}_{y} 
    \end{pmatrix} .
\end{align}
The above results are valid for a system with infinite (or, equivalently, zero) aspect ratio $\kappa = q_{\beta} / q_{\alpha}$. Since the ${\bf q} \to 0$ limit is path dependent, i.e., it depends on the direction of approach towards the origin in the ${\bf q}-$space. We can specify an arbitrary direction through a unit vector ${\hat {\bf n}} \equiv \{ {\hat n}_{\alpha} = \frac{q_{\alpha}}{q} \}$ with ${\alpha } \in [1, 2,  \dots, d]$ and $q = \sqrt{ \sum_{\alpha} q_{\alpha}^2}$.
Now, in the small-${\bf q}$ limit, the structure factor can be written, in the leading order, as
\begin{align}
    S(\textbf{q}) \simeq  \frac{\sum_{\alpha=1}^d \gamma_0^\alpha q_\alpha^2}{2\sum_{\alpha=1}^d D_{\alpha\alpha}q_\alpha^2} = \frac{\textbf{q}^\mathsf{T}\mathbf{\Gamma}^{(0)}\textbf{q}}{2\textbf{q}^\mathsf{T}\textbf{D}\textbf{q}} = \frac{\textbf{q}^\mathsf{T}\mathbf{\chi} \textbf{q}}{\textbf{q}^\mathsf{T} \textbf{D}\textbf{q}}
\end{align}
where we represent the wave vectors as $\textbf{q} = |q_1, q_2, \dots, q_d \rangle$ and $\textbf{q}^T = \langle q_1, q_2, \dots, q_d  |$ and the mobility tensors $\chi$ and $\mathbf{\Gamma}^{(0)}=\mathbf{\Gamma}(\textbf{q}=0)$ at the limit of zero wave number. For the models considered in this work, if there is a biasing field in a particular direction, there is no response in the net current in the orthogonal directions. Thus, the diffusion $\textbf{D}$ and mobility $\mathbf{\Gamma}$ tensors are diagonal.
Furthermore, now taking the limit ${\bf q} \to 0$ in the direction along unit vector $\hat{{\bf n}}$, we can write a ``path-dependent'' (or, aspect-ratio dependent) structure factor $S(\textbf{q}\to 0|\hat{{\bf n}}) \equiv \mathcal{S}^{(0)}(\hat{\textbf{n}})$ as given below:
\begin{align}
\label{FDR-path}
      \mathcal{S}^{(0)}(\hat{\textbf{n}}) = \frac{\hat{\textbf{n}}^\mathsf{T} \mathbf{\Gamma}^{(0)} \hat{\textbf{n}}}{2\hat{\textbf{n}}^\mathsf{T} \textbf{D} \hat{\textbf{n}}} = \frac{\text{Tr}(\mathbf{\Gamma}^{(0)} \mathbf{P}_{\hat{\mathbf{n}}})}{2 \text{Tr}(\mathbf{D} \mathbf{P}_{\hat{\mathbf{n}}})} = \frac{\text{Tr}(\mathbf{\chi} \mathbf{P}_{\hat{\mathbf{n}}})}{\text{Tr}(\mathbf{D} \mathbf{P}_{\hat{\mathbf{n}}})}
\end{align}
where $\mathbf{P}_{\hat{\mathbf{n}}}=\hat{\mathbf{n}} \hat{\mathbf{n}}^\mathsf{T}$ is the projection operator. The last two equalities are obtained by using that the mobility and the bulk-diffusion tensors, $\mathbf{\Gamma}^{(0)}$, ${\bf \chi}$ and ${\bf D}$, respectively, all are diagonal for the class of systems considered here. 
Specifically, in two dimensions, the unit vector $\hat{\bf n} = | \cos \theta, \sin \theta \rangle$ can be parametrized through an angle $\theta$, which is the polar angle in the ${\bf q}-$plane. Then, we can write the density fluctuation (or, the structure factor as ${\bf q} \to 0$) in terms of the bulk-diffusion coefficient and  mobility,
\begin{align}
    \mathcal{S}^{(0)}(\theta) = \frac{\chi^{11} \cos^2 \theta + \chi^{22} \sin^2 \theta}{(D_{11} \cos^2 \theta + D_{22} \sin^2 \theta)} = \frac{\chi^{11} + \kappa^2 \chi^{22}}{(D_{11} + \kappa^2 D_{22}) } ,
\end{align}
where the aspect ratio of a (rectangular) system is $\kappa = q_y/q_x=\tan \theta$.

\begin{figure*}
    \centering
    \includegraphics[width=0.48\linewidth]{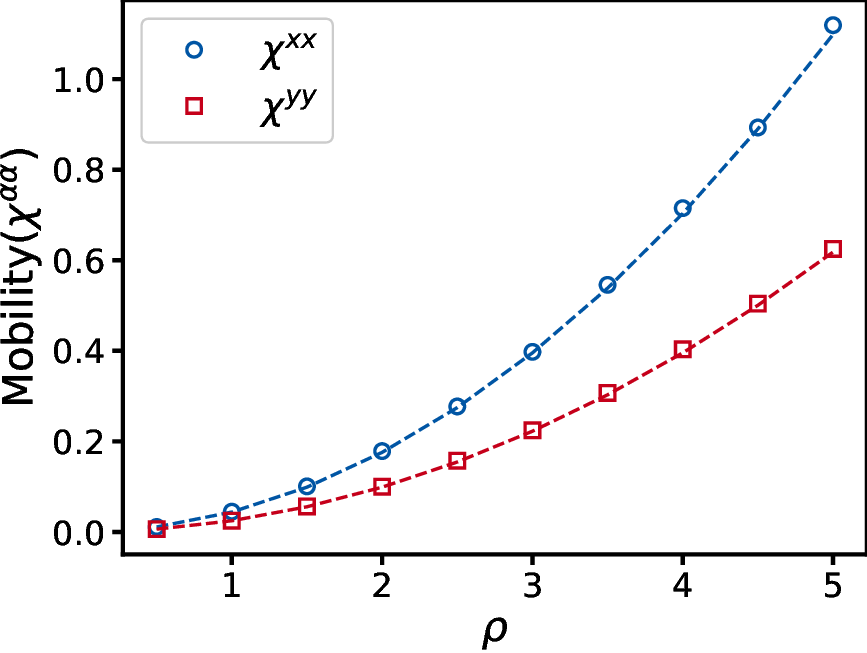}
    \put(-85,120){(a)}
    \includegraphics[width=0.48\linewidth]{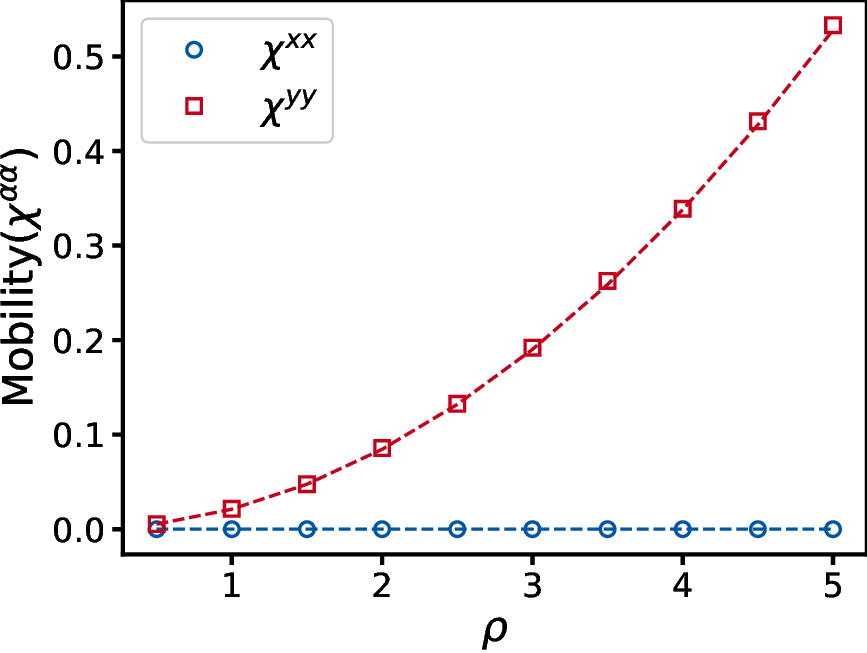}
    \put(-85,120){(b)}
    \caption{The macroscopic (hydrodynamic) mobility is plotted  as a function of density for (a) MCM I and (b) CoMC IB. The symbols represent simulations data, while the dashed lines (in the corresponding colors) denote the analytical predictions for the respective models, as given in Eqs.~\eqref{eq:FDR-mobility-alpha}--\eqref{eq:FDR-mobility-beta} and Table~\ref{tab:mobility_tensor}. The simulations are performed on a system of size $L = 150 \times 150$ with total simulation time $T = 400$. In all panels, the chipping parameters were fixed at $\zeta_1 = 0.2$ and $\zeta_2 = 0.4$. }
    \label{fig:mobility}
\end{figure*}

This nonequilibrium ``fluctuation–dissipation'' relation plays a central role in determining the nature of long-range correlations in the system. In particular, the presence or absence of directional asymmetry in the structure factor around the origin (${\bf q}=0$) directly governs whether, and what kind of, algebraic density correlations emerge in a nonequilibrium steady state. These aspects are summarized in Table~\ref{tab:mobility_tensor} in the models studied in the present work.
In Fig.~\ref{fig:mobility}, we plot the diagonal components of the mobility tensor, $\chi^{\alpha\alpha}$, as a function of the global density for two anisotropic models: (a) MCM I and (b) CoMC IB. The symbols represent simulation results obtained from the scaled, time-integrated total bond current, as defined in Eq.~\eqref{curr-fluc-mob}. The dashed lines correspond to the analytical prediction $\gamma_0^\alpha/2$, derived from the small-$\mathbf{q}$ limit of the structure factor for the respective model with the chosen parameters. One can see an excellent agreement between the simulation data and the theoretical predictions.

\section{Discussion}
\label{sec:discussion}

In this work, we have presented exact calculations of steady-state density–correlation functions in a broad class of conserved-mass transport processes defined on a $d$-dimensional hypercubic lattice with anisotropic, reflection-symmetric hopping. 
The dynamics conserves total mass and, depending on the variant considered, may additionally conserve the center of mass (CoM) along either selected spatial axes or all principal axes. This setting enables a systematic investigation of how anisotropy and conservation laws shape large-scale density fluctuations.

The anisotropy enters through direction-dependent hopping rates that preserve reflection symmetry while breaking rotational invariance. Importantly, all models considered here belong to the class of ``gradient''-type systems \cite{Bertini2001Jul, Arita2014Nov}, with the bulk-diffusion coefficients that are independent of the density. This particular property leads to a significant simplification in the calculations: The hierarchy of correlation functions closes at each order, i.e., there is no BBGKY-type hierarchy involved among correlation functions of different orders. As a consequence, the correlations can be computed exactly, even though they can display nontrivial algebraic scaling.

We have characterized spatial correlations through the structure factor and identified the conditions under which algebraic decay and hyperuniformity emerge in such systems.
In the absence of CoM conservation, anisotropic mass-conserving systems are known to exhibit long-ranged correlations with power-law decay of the form
$C(\mathbf{x}) \sim {1}/{|\mathbf{x}|^d},$
in agreement with earlier studies \cite{Garrido1990Aug, Maes1990Nov}. This $1/|\mathbf{x}|^d$ scaling is a hallmark of anisotropic systems with mass-conserving dynamics, and reflects the emergence of effective quadrupolar structures of the Poisson equation governing the steady-state correlation functions.

To incorporate CoM conservation, we have introduced coordinated multidirectional hopping events in which equal mass chunks move simultaneously in opposite directions. These paired moves violate detailed balance and thus drive the system to a genuine nonequilibrium non-Gibbsian steady state, distinct from an equilibrium one described by the familiar Boltzmann–Gibbs distribution. While preserving total mass, they also conserve the center of mass of the transported mass. Depending on whether CoM conservation is enforced along some particular (not all) directions or along all directions, the systems exhibit qualitatively distinct scaling behaviors.

When CoM is conserved along {\it all} principal axes, the steady-state correlation functions decay substantially faster, typically as
$C(\mathbf{x}) \sim {1}/{|\mathbf{x}|^{d+2}},$
{\it irrespective of whether the microscopic hopping rates are isotropic or anisotropic.} Thus, the presence of anisotropy fails to restore the slower $1/|\mathbf{x}|^d$ decay once full CoM conservation is enforced. Instead, the additional conservation law fundamentally alters the structure of fluctuations. Although the decay remains algebraic, the faster $1/|\mathbf{x}|^{d+2}$ scaling implies a significant suppression of long-wavelength density fluctuations. In Fourier space, this corresponds to a structure factor that vanishes anomalously rapidly as wave number ${\bf q} \to {\bf 0}$, placing the system in the category of an extreme, “class I” {\it hyperuniform} states. In such states, density fluctuations at large scales are suppressed more strongly than in typical disordered systems, despite the absence of crystalline order.

By contrast, when CoM conservation is imposed only along some particular direction(s) (while mass remains conserved globally), the slower $1/|\mathbf{x}|^d$ decay is recovered. In this case, the constraint is insufficient to eliminate the dominant multipolar contribution responsible for the long-range correlations characteristic of anisotropic mass-conserving dynamics. Partial CoM conservation therefore does not qualitatively alter the leading-order scaling of correlations, although it can modify anisotropic angular dependence of correlations.

The origin of these distinct power-law scaling can be understood through an analogy with electrostatics. The steady-state correlation function satisfies an effective Poisson-type equation in which the conserved quantities determine the multi-polar structure of an emergent ``charge'' distribution. When only mass is conserved, the leading non-vanishing contribution in the multipole expansion corresponds to a (rank-2) quadrupolar charge distribution, yielding a potential that decays as $1/|\mathbf{x}|^d$; both monopolar and dipolar contributions vanish by mass conservation and reflection symmetry. However, when CoM is conserved in all directions,  the quadrupolar term is also suppressed by the additional constraints and the corresponding prefactor vanishes. The leading contribution then arises from a higher-order (rank-4) multi-polar charge distribution, which produces the faster $1/|\mathbf{x}|^{d+2}$ decay of the density correlation function.

Furthermore, for the class of models considered here, we have established a nonequilibrium fluctuation–dissipation relation [Eq. \eqref{FDR-path}] that connects the macroscopic mobility, bulk-diffusion coefficients, and static structure factor. A salient feature of such anisotropic systems is that, in dimensions higher than one, the small–wave-number limit of the structure factor can depend on the path of approach in reciprocal space. By taking this limit along different directions (say, along a principal axis), one can define a ``path-dependent'' structure factor that characterizes long-wavelength density fluctuations where (scaled) mass or particle-number fluctuations within a large subsystem depend on the subsystem aspect ratio (or, shape).
We showed that the macroscopic mobility tensor can be expressed in terms of current fluctuations in the hydrodynamic limit. In particular, the diagonal elements of the macroscopic mobility tensor are proportional to the variance of the total space–time-integrated current across the whole system, scaled by the space–time volume. This provides a direct connection between current fluctuations and the hydrodynamic transport coefficients. 
Moreover, in the absence of cross diffusion (i.e., force along a particular direction does not induce currents along its orthogonal directions), the bulk-diffusion tensor remains diagonal in these anisotropic systems, and subsequently the off-diagonal elements of the macroscopic mobility tensor vanish.

\section{Summary and concluding remarks}
\label{sec:conclusion}

In summary, we have developed a theoretical framework to systematically investigate the interplay between anisotropy and center-of-mass (CoM) conservation and to determine how their competition shapes the spatial structure of fluctuations in nonequilibrium steady states. Our analysis demonstrates that the mere presence of microscopic anisotropy does not automatically guarantee 
$1/r^d$ scaling of the density correlation function. Rather, the nature of the conservation laws—and the precise manner in which anisotropy enters the dynamics—plays a crucial role in determining the form of the steady-state correlations.

Power-law correlations are a generic feature of nonequilibrium systems with conserved quantities, even far from criticality. In anisotropic systems with inversion symmetry and a single conserved scalar field (such as mass or particle number), a singularity in the structure factor at small wave number gives rise to algebraically decaying correlations whose scaling is dictated by symmetry and dimensionality. However, we have shown that the inclusion of an additional CoM conservation law fundamentally modifies the structure of this singularity and can thereby qualitatively alter the resulting spatial correlations.

Specifically, while anisotropy combined with a single conservation law typically produces long-range correlations decaying as $1/r^d$, the imposition of full CoM conservation leads to a stronger suppression of long-wavelength density fluctuations and changes the decay to $1/r^{d+2}$, resulting in hyperuniform behavior. In contrast, partial CoM conservation—implemented only along selected directions—does not modify the decay exponent, although it can significantly affect the amplitudes and even the signs of the correlations along different axes. Thus, the competition between anisotropy and CoM conservation gives rise to a rich phenomenology of steady-state spatial structures in a broad class of nonequilibrium systems with multidirectional hopping.

Overall, our results provide a unified microscopic approach for understanding how competing mechanisms—spatial symmetries (or, lack of them) and conservation laws—govern the emergence of large-scale structures in diffusive systems that violate detailed balance. 
Indeed, they highlight how conservation laws beyond simple mass conservation qualitatively reshape long-range correlations in nonequilibrium steady states. In particular, they demonstrate that the multipolar structure imposed by microscopic conservation principles directly governs the universal large-distance behavior of correlation functions.
We expect that the framework developed here will be applicable to a wide range of interacting-particle systems and may offer further insight into anomalous fluctuations and hyperuniformity observed in driven systems in general.

\section*{Acknowledgments} 

We thank Tanmoy Chakraborty and Arghya Das for reading the manuscript. 

%XXXXXXX

\appendix
%\widetext
\section{Mobility and current fluctuation}\label{app:mob_cur}

Here we derive the relation between the {\it macroscopic} (space-averaged) mobility and the variance of space-time-integrated current across the whole system.
We begin with the time-evolution equation for the time-integrated current, which, in a diffusive system, takes the form,
\begin{align}\label{eq:curqt}
    \frac{d}{dt}\langle \mathcal{Q}_\alpha(\mathbf{r}, t)\rangle 
    = \langle {\cal J}^{({\rm d})}_{\alpha}(\mathbf{r}, t) \rangle,
\end{align}
where the right-hand side represents the average diffusive component of the instantaneous bond current. The total bond current can be decomposed as
\begin{align}\label{eq:cur}
    \mathcal{J}_\alpha(\mathbf{r}, t) 
    = {\cal J}^{({\rm d})}_{\alpha}(\mathbf{r}, t) 
    + {\cal J}^{({\rm fl})}_{\alpha}(\mathbf{r}, t),
\end{align}
with the average $\langle {\cal J}^{({\rm fl})}_{\alpha}(\mathbf{r}, t)\rangle = 0$ of the fluctuating (``noise'') current component vanishes by definition. The precise form of the fluctuating current component depends on the microscopic details of the model and has been computed explicitly for the systems considered here (see main text).
We now consider the equal-time two-point connected correlation function of the time-integrated bond current,
\[
C_{\mathbf{r}}^{\mathcal{Q}_\alpha \mathcal{Q}_\beta}(t)
= \langle \mathcal{Q}_\alpha(\mathbf{0}, t)
  \mathcal{Q}_\beta(\mathbf{r}, t)\rangle_c.
\]
Its time evolution is given by
\begin{align}\label{eq:q0qrgammar}
\frac{d}{dt} 
C_{\mathbf{r}}^{\mathcal{Q}_\alpha \mathcal{Q}_\beta}(t,t)
= C_{\mathbf{r}}^{j^{({\rm d})}_\alpha \mathcal{Q}_\beta}(t,t)
+ C_{\mathbf{r}}^{\mathcal{Q}_\alpha j^{({\rm d})}_\beta}(t,t) + \Gamma_{\mathbf{r}}^{\alpha\beta},
\end{align}
where the quantity $\Gamma_{\mathbf{r}}^{\alpha \beta}$ is ``microscopic'' ({\it space-dependent}) mobility tensor introduced in the main text.  
Now, using Eqs.~\eqref{eq:curqt} and \eqref{eq:cur}, the left-hand side can be written explicitly as
\begin{align}\label{eq:q0qr} 
\frac{d}{dt} C_\textbf{r}^{\cQ_\alpha \cQ_\beta}(t, t) = \langle \left( {\cal J}^{({\rm d})}_\alpha(\textbf{0}, t) + {\cal J}_\alpha^{({\rm fl})}(\textbf{r}, t) \right) \cQ_\beta(\textbf{r}, t) \rangle_c \notag 
\\
+ \langle \cQ_\alpha(\textbf{0}) \left( {\cal J}^{({\rm d})}_\beta(\textbf{r}, t) + {\cal J}_\beta^{({\rm fl})}{(\textbf{r} ,t)} \right) \rangle_c. 
\end{align}
Substituting Eq.~\eqref{eq:q0qr} into Eq.~\eqref{eq:q0qrgammar}, we obtain
\begin{align}
\label{eq:q0qrjfl} 
\langle {\cal J}_\alpha^{({\rm fl})}(\textbf{0}, t) \cQ_\beta(\textbf{r},t) \rangle_c + \langle \cQ_\alpha(\textbf{0},t) {\cal J}_\beta^{({\rm fl})}(\textbf{r}, t) \rangle_c = \Gamma_{\textbf{r}}^{\alpha \beta}. 
\end{align}
Using the definition of the time-integrated current,
$\cQ_\alpha(\textbf{r},t)=\int_0^t dt' [{\cal J}^{(d)}_{\alpha}(\textbf{r}, t') + {\cal J}^{({\rm fl})}_{\alpha}(\textbf{r}, t')]$,
Eq.~\eqref{eq:q0qrjfl} becomes
\begin{align} 
&\int_0^t dt'\left[\langle {\cal J}_\alpha^{({\rm fl})}(\textbf{0}, t) {\cal J}^{({\rm fl})}_{\beta}(\textbf{r}, t') \rangle_c + \langle {\cal J}^{({\rm fl})}_{\alpha}(\textbf{0}, t') {\cal J}_\beta^{({\rm fl})}(\textbf{r}, t) \rangle_c \right]\notag\\&= \Gamma_{\textbf{r}}^{\alpha \beta}, \end{align}
where we have taken $\langle 
{\cal J}_\alpha^{({\rm fl})}(\mathbf{0}, t) 
{\cal J}^{({\rm d})}_\beta(\mathbf{r}, t') 
\rangle = 0$ for $t > t'$.
Rewriting the above equation by rearranging the limits of integration, we get 
\begin{align}
\int_{-t}^{t} dt' \langle {\cal J}_\alpha^{({\rm fl})}(\mathbf{0}, t)  {\cal J}_\beta^{({\rm fl})}(\mathbf{r}, t') 
\rangle =\Gamma_{\mathbf{r}}^{\alpha\beta},
\end{align}
which holds for any time $t$ and therefore implies the (two-point) fluctuating- or `noise''-current statistics,
\begin{align}
    \langle {\cal J}_\alpha^{({\rm fl})}(\mathbf{0}, t) 
{\cal J}_\beta^{({\rm fl})}(\mathbf{r}, t) 
\rangle =\Gamma_{\mathbf{r}}^{\alpha\beta} \delta(t).
\end{align}
That is, the space-time correlation functions for fluctuating current (``noise'') are delta-correlated in time, but have a finite short-ranged correlation in space.

{\it The macroscopic mobility:} We define the (scalar) macroscopic or hydrodynamic mobility, which is the scaled variance of space-time-integrated current, say, along $\alpha$ direction, in the thermodynamic limit as 
\begin{align}
\nonumber\chi^{\alpha\alpha}&=\lim_{T\to \infty, L\to \infty}\frac{1}{2 TL^d}\left\langle\left[\sum_{\textbf{r}}\cQ_{\alpha}(\textbf{r}, T)\right]^2\right\rangle 
\\ 
&=\lim_{T\to \infty, L\to \infty}\frac{1}{2 TL^d}\left\langle\left[\sum_{\textbf{r}}\cQ_{\alpha}^{({\rm fl})}(\textbf{r}, T)\right]^2\right\rangle.
\end{align}
The above equation can be rewritten in terms of correlations between the instantaneous fluctuating current at two space and time points as follows:
\begin{align}
    \chi^{\alpha\alpha} &= \lim_{T\to \infty, L \to \infty}\frac{1}{2 TL^d} 
    \\
    &\times \sum_{\textbf{r}, \textbf{r}'} \int_0^Tdt \int_0^T dt' \langle {\cal J}_\alpha^{({\rm fl})}(\textbf{r}, t) {\cal J}_\alpha^{({\rm fl})}(\textbf{r}', t') \rangle \\
    &=\frac{1}{2 } \sum_{\textbf{r}}\int_{-\infty}^\infty dt C^{{\cal J}_\alpha^{({\rm fl})} {\cal J}_\beta^{({\rm fl})}}(t, 0) = \frac{1}{2} \sum_{\textbf{r}} \Gamma^{\alpha \alpha}_{\textbf{r}},
\end{align}
which leads to the desired relation Eq. \eqref{curr-fluc-mob} discussed in the main text.

%\section{Fluctuation-dissipation relation (FDR)}

\section{Anisotropic Green's function}
\label{app:Green}

In this section, we provide details of the scheme to calculate the long-range contributions to the correlation function given in Eq.~\eqref{corr}. Specifically, we aim to evaluate the following integrals (inverse Fourier transforms),
\begin{equation}\label{1/r^d}
    C_{A1}({\bf r}) = \int_{BZ} \frac{d^{d} {\bf q}}{(2\pi)^{d}}  \frac{\sum_{\alpha}\gamma_{0}^{\alpha}\lambda(q_{\alpha})}{{\sum_\alpha 2D_{\alpha\alpha} \lambda(q_\alpha)}} e^{-i{\bf q.r}},
\end{equation}
and
\begin{equation}
    C_{A3}({\bf r}) = \int_{BZ}\frac{d^{d} {\bf q} }{(2\pi)^{d}}\frac{\sum_{\alpha}(\gamma_{1}^{\alpha}-(\gamma_{2}^{\alpha})^{2})\lambda^{2}(q_{\alpha})}{{\sum_\alpha 2D_{\alpha\alpha} \lambda(q_\alpha)}}e^{-i{\bf q}.{\bf r}}.
\end{equation}
We can rewrite the above equations as given below:
\begin{equation}\label{eq:CA1}
C_{A1}({\bf r})\simeq-\sum_{\alpha}\gamma_{0}^{\alpha}\frac{\partial^2}{\partial x_{\alpha}^2}G_{A}(r),
\end{equation}
and
\begin{align}\label{eq:CA3}
    C_{A3}({\bf r}) &\simeq \sum_{\alpha} F_{\alpha}\frac{\partial^4}{\partial x_{\alpha}^4} G_{A}(r),
\end{align}
where the prefactor  is given by $F_{\alpha} \equiv \gamma_{1}^{\alpha}-(\gamma_{2}^{\alpha})^{2}$
and the Green's function,
\begin{equation}
    G_{A}({\bf r})=\int_{BZ}\frac{d^{d}q}{(2\pi)^{d}}\frac{1}{{\sum_{\alpha=1
    } ^{d} 2D_{\alpha\alpha}\lambda(q_\alpha)}}e^{-iq.r}.
\end{equation}
We then use the identity $1/A=\int_{0}^{\infty } e^{-At}dt$ to write the Green's function as 
\begin{equation}
    G_{A}({\bf r})=\int_{BZ}\frac{d^{d}q}{(2\pi)^{d}}\int_{0}^{\infty}dt. e^{-\sum_\alpha 2D_{\alpha\alpha}\lambda(q_\alpha)t}e^{-iq.r}.
\end{equation}
Since we are interested in small-${\bf q}$ (large-${\bf r}$) asymptotics, we approximate $\cos{q_{\alpha}} \simeq (1 - q_{\alpha}^{2}/2)$ and write the Green's function,
\begin{align}\label{eq:greens_fnc}
    G_{A}({\bf r})&= \int_{0}^{\infty}dt\int_{BZ}\frac{d^dq}{(2\pi)^d} e^{\sum_{\alpha}(-2tD_{\alpha\alpha}q_{\alpha}^{2}-iq_{\alpha}x_{\alpha})}
    \notag\\
&= \int_{0}^{\infty}dt \left( \prod_{\alpha}\sqrt{\frac{1}{8\pi D_{\alpha\alpha}t}} \right) e^{-\sum_{\alpha} {x_{\alpha}^{2}}/{8D_{\alpha\alpha}t}} . 
\end{align}
First, we define the rescaled coordinates to handle the anisotropy $D_{\alpha\alpha}$. We define the rescaled position vector $\mathbf{R}$ with components $X_\alpha$:
$X_\alpha = x_\alpha/\sqrt{D_{\alpha\alpha}}$
The magnitude is given by $R = \sqrt{\sum_\alpha X_\alpha^2}$.
Substituting the above expression for the Green's function \eqref{eq:greens_fnc} into Eqs.~\eqref{eq:CA1} and \eqref{eq:CA3}, we obtain
\begin{align}\label{eq:CrA1_asym}
    C_{A1}(\textbf{r}) \simeq
    \frac{\Gamma(d/2)}{4 \pi^{d/2} \sqrt{\prod D_{\mu\mu}}} \sum_{\alpha=1}^d \frac{\gamma_0^\alpha}{D_{\alpha\alpha}} \left[ 1 - d \left( \frac{X_\alpha}{R}\right)^2 \right]\frac{1}{R^d},
\end{align}
and 
%\begin{widetext}
\begin{align}\label{eq:CrA3_asym}
    C_{A3}({\bf r}) 
    \simeq\frac{3 \Gamma(d/2 + 2)}{(d+2)\pi^{d/2} \sqrt{\prod D_{\mu\mu}}} \frac{1}{R^{d+2}} \sum_{\alpha=1}^d \frac{\gamma_1^\alpha - (\gamma_2^\alpha)^2}{D_{\alpha\alpha}^2}\times \notag\\ \left[ 1 - 2(d+2)\left(\frac{X_\alpha}{R}\right)^2 + \frac{(d+2)(d+4)}{3}\left(\frac{X_\alpha}{R}\right)^4 \right].
\end{align}
%\end{widetext}
Here, we have used the identity $\int_{0} ^{\infty} x^{-b} e^{-a/x}dx=a^{1-b}\Gamma(b-1)$, which is valid for ${\rm Re}(a)>0 $ and ${\rm Re(b)}>1$. It is worth noting that Eqs.~\eqref{eq:CrA1_asym} and \eqref{eq:CrA3_asym} exhibit leading-order decay as $1/R^{d}$ and $1/R^{d+2}$, respectively. In addition, subleading corrections are present: Eq.~\eqref{eq:CrA1_asym} contains terms of order $\mathcal{O}(1/R^{d+2})$, while Eq.~\eqref{eq:CrA3_asym} includes terms of order $\mathcal{O}(1/R^{d+4})$. In this work, we have focused on the leading-order terms (large-scale asymptotic) in the correlation functions, and the higher-order power-law contributions have been neglected. But, these higher-order corrections can be systematically obtained through small$-{\bf q}$ expansion of the structure factor and will be addressed elsewhere.

\twocolumngrid

%% Reference to your .bib file
\bibliographystyle{apsrev4-2}
\bibliography{references}

\end{document}